\documentclass[
                trackchanges, twocolumn]{aastex701}

\usepackage{
  graphicx,
  bm,
  amsmath,
  amssymb,
  mathtools,
  tikz,
  tikz-cd,
  mathdots,
  minted,
  tabularx,
  booktabs,
adjustbox}
\usepackage[dvipsnames]{xcolor}

\newcommand{\entity}{\texttt{Entity}}

\newcommand{\zz}{\texttt{Zigzag} }
\newcommand{\es}{\texttt{Esirkepov} }
\newcommand{\trd}{$3^\mathrm{rd}$}
\newcommand{\rd}{$^\mathrm{rd}$}
\newcommand{\st}{$^\mathrm{st}$}
\newcommand{\thu}{$^\mathrm{th}$}
\newcommand{\nd}{$^\mathrm{nd}$}
\newcommand{\mS}{\mathcal{S}}

\newcommand{\mW}{\mathcal{W}}

\makeatletter
\newcommand{\subalign}[1]{%
  \vcenter{%
    \Let@ \restore@math@cr \default@tag
    \baselineskip\fontdimen10 \scriptfont\tw@
    \advance\baselineskip\fontdimen12 \scriptfont\tw@
    \lineskip\thr@@\fontdimen8 \scriptfont\thr@@
    \lineskiplimit\lineskip
    \ialign{\hfil$\m@th\scriptstyle##$&$\m@th\scriptstyle{}##$\hfil\crcr
      #1\crcr
    }%
  }%
}
\makeatother

\begin{document}

\title{\entity~-- Hardware-agnostic Particle-in-Cell Code for Plasma Astrophysics.\\III: Higher-order shape functions \& generalized field stencils}
\correspondingauthor{Ludwig M. B\"oss}

\author[orcid=0000-0003-4690-2774]{Ludwig M. B\"oss}
\affiliation{Department of Astronomy and Astrophysics, The University of Chicago, William Eckhart Research Center, 5640 S. Ellis Ave. Chicago, IL 60637, USA}
\email[show]{lboess@uchicago.edu}

\author[orcid=0000-0002-3643-9205]{Arno Vanthieghem}
\affiliation{Sorbonne Université, Observatoire de Paris, Université PSL, CNRS, LUX, F-75005 Paris, France}
\email{}

\author[orcid=0000-0001-8939-6862]{Hayk Hakobyan}
\affiliation{Center for Computational Astrophysics, Flatiron Institute, New York, NY 10010, USA}
\email{}

\author[orcid=0000-0001-9073-8591]{Evgeny A.  Gorbunov}
\affiliation{Department of Physics, University of Maryland, College Park, MD 20742, USA}
\email{}

\author[orcid=0000-0003-4690-2774]{Damiano Caprioli}
\affiliation{Department of Astronomy and Astrophysics, The University of Chicago, William Eckhart Research Center, 5640 S. Ellis Ave. Chicago, IL 60637, USA}
\email[]{}

\begin{abstract}

  Modern particle-in-cell (PIC) codes have become an integral tool in plasma astrophysics.
  As most plasma phenomena grow from initially small instabilities, it is important to ensure PIC codes can suppress noise and ensure that any growing instability is indeed physical.
  Therefore, we introduce our efforts to implement higher-order methods for the current deposit and field interpolation as well as generalized field stencils for the field solver in the PIC code \entity.
  Our updated current deposit scheme allows for up to 11\thu-order accurate interpolation, while the generalized stencils for the field solver can be tuned to suppress numerical dispersion.
  We perform extensive tests to ensure high accuracy of the implemented schemes for charge conservation, stabilization against numerical heating, improved energy conservation, and suppression of numerical Cherenkov effects. To supply a benchmark on performance impact, we demonstrate the scaling of the higher-order current deposit and discuss the possible performance balance between higher-order interpolation, numerical resolution, and the inclusion of additional current filtering.

\end{abstract}


\section{Introduction\label{sec:intro}}
Particle-in-cell (PIC) simulations have become a powerful tool to improve our understanding of astrophysical plasma processes \citep[e.g.][for reviews]{Verboncoeur2005, Arber2015, Pohl2020, Marcowith2020, Nishikawa2021}.
These simulations explicitly solve the Vlasov-Maxwell set of equations and can therefore self-consistently model complex plasma-kinetic systems.

In the context of plasma-astrophysics, PIC simulations allow to make predictions for the dynamics of plasma turbulence \citep[e.g.][]{Zhdankin2017b, Zhdankin_2018,Zhdankin2017, Zhdankin2019, Comisso2018, Comisso2019,Comisso_2022,gorbunov_2025b}, particle acceleration at shocks in various environments \citep[][]{Spitkovsky2008, Spitkovsky2008a, Sironi2011, Sironi2013, Park2015, Bohdan2017, Ha2018, Lemoine2019, Ha2021, Kobzar2021, Cerutti2023, Gupta2024, Vanthieghem2024, Groselj2024}, cosmic ray driven instabilities \citep[e.g.][]{Niemiec2008, Riquelme2009, Ohira2009, Gargate2010, Holcomb2019, Shalaby2021, Zhou2024, Lemmerz2025}, thermal conduction \citep[e.g.][]{Roberg-Clark2018, Komarov2018}, magnetic reconnection \citep[e.g.][for a review]{Sironi.etal_2025}, pulsar magnetospheres \citep[e.g.][]{Philippov.Spitkovsky_2014,Chen.Beloborodov_2014,Bransgrove.etal_2023,Hakobyan2023,Cerutti.etal_2020,Philippov.Kramer_2022}, and black hole accretion disks and coronae \citep[e.g.][]{Riquelme_2012,Hoshino_2015,Parfrey_2019,Bacchini_2022,Galishnikova.etal_2023,Bacchini_2024,Groselj_2024,Nattila2024,sandoval_2024,Gorbunov2025, Groselj2026}

All of these studies are contingent on numerical convergence to distinguish physical effects from numerical noise and other artificial processes, while being constrained by computational resources. Convergence can typically be achieved by either increasing the numerical resolution, i.e., the number of cells or the number of particles per plasma skin-depth, or improving the accuracy of the solver.

In PIC simulations, the coupled evolution of particles and fields is ensured through successive steps: electric currents from particles are deposited onto the grid to compute the current or charge density, while the electromagnetic fields are subsequently interpolated back to particle positions to advance their trajectories. Macroparticles in PIC represent numerical characteristics of the Vlasov equation, discretely sampling the distribution function: $f(\bm{x}, \bm{u}) \approx \sum_p \mathcal{S}(\bm{x} - \bm{x}_p)\delta(\bm{u}-\bm{u}_p)$, where $\bm{x}_p$ and $\bm{u}_p$ are the coordinates and velocities associated with the $p$-th particle, and $\mathcal{S}$ is the so-called shape function (finite in extent and normalized to unity). 
Since electromagnetic PIC algorithms typically avoid solving the expensive Poisson's equation ($\nabla\cdot\bm{E}=4\pi\rho$) on the grid, 
a special class of current deposition schemes has been developed to explicitly satisfy charge conservation from the continuity equation, $\partial_t\rho + \nabla\cdot\bm{J}=0$.
For the simple zero-th and first order polynomial shape functions, the algorithms developed by \citet{Villasenor1992} and \citet{Umeda2003} are the most commonly used, while for the general case of shape functions using polynomials of arbitrary order, the most widely used implementation is the one developed by \citet{Esirkepov2001}.

Crucially, to properly solve Vlasov's equation with particle characteristics, the positions associated with the shape functions must be updated with the interpolated fields convolved using the same shape function kernel: $\{\bm{E}_p,\bm{B}_p\} = \int d^3 \bm{x} \{\bm{E},\bm{B}\}\mathcal{S}(\bm{x}-\bm{x}_p)$ which adds to the overall performance penalty. Thus, employment of various shape function orders represents a trade-off between the speed of low-order and the reduced aliasing noise of high-order interpolation.

Numerical noise typically manifests in artificial heating of the particle distribution, leading to spurious acceleration and back-reaction with the electric and magnetic field. The inclusion of current filtering \citep{Vay2011} can alleviate some of the numerical noise introduced by using low-order interpolation in the current deposit.\footnote{Currents are typically filtered using the so-called digital filter, smoothing the deposited currents in multiple passes with a finite-window $(1/4,1/2,1/4)$ stencil to approximate a Gaussian blur in the limit of infinite filter passes, and with that effectively increasing the order of the shape function used for the current deposit.} This, however, comes at a cost of potentially smoothing out otherwise physical gradients in the current densities.

The use of higher-order shape functions for constructing the current density is a well-established method \citep[e.g.][for a review]{Arber2015} and has since been employed in various modern PIC codes, such as \textsc{Osiris} \citep{Fonseca2003, Fonseca2018}, \textsc{Smilei} \citep{Derouillat2018}, and \textsc{Full EPIC-GOD} \citep{Kim2026}.
Similarly, their use to construct the charge density grid in Poisson-solver-based PIC codes has been shown to improve numerical stability considerably \citep[][]{Shalaby2017}.

Another key aspect in solving the Vlasov-Maxwell system is the choice of finite derivatives, specifically when evolving the magnetic field using Faraday's law: $\partial_t \bm{B}=-c\nabla\times\bm{E}$. Since in typical finite-difference time domain (FDTD) Maxwell-solvers, the fields on the grid are staggered according to the classical \cite{Yee1966} mesh convention, the simplest approach is to construct finite differences using only two adjacent grid-points. This method can be extended to use more adjacent grid-points for the difference stencil, with the option to optimize these stencils for specific dispersion relations \citep[e.g.,][]{Pukhov1999, Lehe2013, Cowan2013, Blinne2018}. These methods have been shown to improve accuracy related to numerical dispersion \citep[e.g.,][]{Cole2002, Karkkainen2006} and mitigate numerical heating due to the numerical Cherenkov instability \citep[e.g.,][]{Cormier-Michel2008, Vay2011, Blinne2018, Lu2020, Filipovic2022, Huddleston2025}.\footnote{Note, that applying filters to the electric fields \citep[typically referred to as ``Friedmann filters'', e.g.,][]{Greenwood2004} has also been shown to mitigate numerical instabilities \citep[][]{Plotnikov2018}.}

Modern, versatile PIC codes have to address the problems of momentum or energy conservation, numerical instabilities, and numerical particle heating, while retaining enough performance to solve state-of-the-art problems in computational plasma physics.

In \citet{Hakobyan2025} we introduced \entity, a performance-portable PIC code, built for massively-parallel problems on state-of-the-art exa-scale GPU clusters, independent of their architecture.
Paper II in this series \citep{Galishnikova2025} introduced the general-relativistic module of \entity.

With this work (Paper III), we introduce higher-order methods for current deposit, field interpolation, and the Maxwell solver, implemented into the latest release of \entity. We discuss the different implementations and their interplay, with particular emphasis on the conditions required to ensure long-term stability and accuracy in astrophysically relevant plasma systems.

This paper is structured as follows: In Sec.~\ref{sec:methods}, we briefly recap the \entity~framework and introduce the changes made for this work; 
In Sec.~\ref{sec:convergence}, we show the impact of these changes on the convergence of various test cases; In Sec. \ref{sec:performance}, we analyze the performance impact of each improvement and define a trade-off metric to give an estimate of when which improvement is worth using; Finally, in Sec.~\ref{sec:conclusion}, we summarize our work.

\section{Methods}
\label{sec:methods}

\subsection{Core PIC algorithm}
\entity~solves a coupled system of Vlasov-Maxwell equations using macroparticle characteristics, which involves integrating the following set of equations:

\begin{align}
  \partial_t\bm{B}&=-c\nabla\times\bm{E},\label{eq:faraday} \\
  \partial_t\bm{E}&=c\nabla\times\bm{B} - 4\pi\bm{J},\label{eq:ampere} \\
  d\bm{u}_p/dt &= (q_s/m_s c)\left(\bm{E}_p+\bm{v}_p\times\bm{B}_p\right), \label{eq:velUpd}\\
  d\bm{x}_p/dt &= c \bm{v}_p, \label{eq:posUpd}
\end{align}

\noindent where the electromagnetic fields, $\bm{E}$ and $\bm{B}$, and the electric current densities, $\bm{J}$, are defined on a discretized staggered Yee-grid, $\bm{u}_p$ and $\bm{x}_p$ are the continuous four-velocity (dimensionless) and position of the $p$-th particle of species $s$ with charge $q_s$ and mass $m_s$, $\bm{v}_p\equiv \bm{u}_p/\sqrt{1+|\bm{u}_p|^2}$, and

\begin{align}
  \label{eq:fieldInterp}
  \{\bm{E}_p,\bm{B}_p\}\equiv \int d^3\bm{x}\mathcal{S}_{\mathcal{O}}(\bm{x}-\bm{x}_p)\{\bm{E},\bm{B}\},
\end{align}

\noindent are the electromagnetic fields interpolated from the grid to the position of the macroparticle using its shape-function $\mathcal{S}_{\mathcal{O}}(\bm{x})$ of order $\mathcal{O}$. The electric currents in \eqref{eq:ampere} are deposited from the macroparticles to satisfy the charge-continuity equation:

\begin{equation}
  \partial_t\rho+\nabla\cdot\bm{J}=0, \label{eq:chargeCont}
\end{equation}

\noindent where $\rho(\bm{x})\equiv dV^{-1}\sum_s\sum_p w_p q_s\mathcal{S}_\mathcal{O}(\bm{x}-\bm{x}_p)$, with $w_p$ being the weight of the $p$-th particle, and $dV$ being the volume of each cell. Note, that correctly solving \eqref{eq:faraday}, \eqref{eq:ampere} on the Yee-mesh, while also conserving \eqref{eq:chargeCont}, automatically ensures that $\nabla\cdot \bm{B}=0$ and $\nabla\cdot\bm{E}=4\pi\rho$ -- which are never explicitly solved by the algorithm.

The description of the time-stepping scheme and the discretization can be found in more detail in Paper I \citep{Hakobyan2025}; here, instead, we focus on the new aspects --- the current deposition, which satisfies \eqref{eq:chargeCont}, and the new discretization used to solve \eqref{eq:faraday}. Note that while the discussion in this paper focuses on Cartesian coordinates, high-order shape functions also work in flat space-time curvilinear and general-relativistic coordinates with very minor adjustments to the scheme (see the appendix~\ref{sec:curvilinear} for a discussion). However, the advanced stencils for the Faraday solver presented in the further section explicitly optimize the dispersion relation globally in Cartesian coordinates, and are thus not easily extensible to a curvilinear basis. 

\subsection{Current deposition with high-order shape functions}
\label{sec:esirkepov}

\begin{figure}
  \centering
  \includegraphics[width=\linewidth]{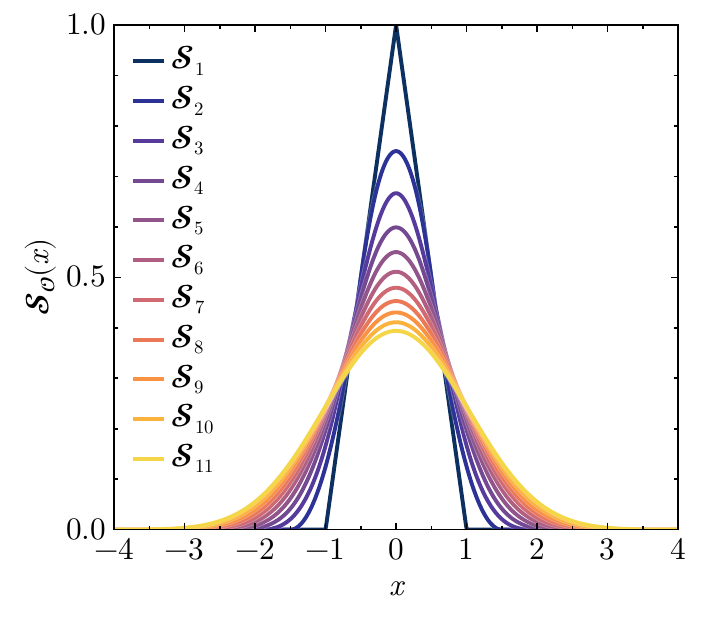}
  \caption{Shape functions $\mathcal{S}_\mathcal{O}$ implemented in \entity, normalized to the grid-cell size $\Delta x$. The subscripts refer to the leading order $\mathcal{O}$ of the polynomial.}
  \label{fig:shape_functions}
\end{figure}
Below, we describe our implementation of the charge-conservative current deposition scheme first proposed by \cite{Esirkepov2001}. While in this section we focus on the flat space-time Cartesian case, the algorithm is exactly the same for a general curvilinear coordinate basis and even in GR, due to the use of intermediate ``conformal'' currents described in more detail in \cite{Hakobyan2025}.

In what follows, we additionally assume for brevity that the sizes of all cells in all directions is $\delta x=\delta y=\delta z = 1$. The first-order shape function in 1D can thus be written as:

\begin{equation}
  \mathcal{S}_1^{\mathtt{1D}}(x) =
  \begin{cases}
    1 - \vert x\vert \quad &, \: \vert x \vert \leq 1
    \\
    0  \quad&, \: \vert x \vert > 1
  \end{cases}
\end{equation}

\noindent In 2D and 3D, the shape function is constructed using a product of individual 1D shape functions in each of the directions, i.e., $\mathcal{S}_\mathcal{O}^{\mathtt{3D}}(\bm{x})=\mathcal{S}_\mathcal{O}^{\mathtt{1D}}(x)\mathcal{S}_\mathcal{O}^{\mathtt{1D}}(y)\mathcal{S}_\mathcal{O}^{\mathtt{1D}}(z)$. For simplicity, we will further drop the $\mathtt{1D}$ notation, assuming: $\mathcal{S}_\mathcal{O}(x)\equiv \mathcal{S}_\mathcal{O}^{\mathtt{1D}}(x)$. General higher-order shape functions can be constructed using piecewise-polynomials or B-splines, while requiring the following conditions:

\begin{enumerate}
  \item symmetry: $S_\mathcal{O}(x)=S_\mathcal{O}(|x|)$;
  \item normalization: $\int S_\mathcal{O}(x)d x=1$;
  \item positivity: $S_\mathcal{O}(x)\geq 0$ for $\forall x$.
  \item finite extent: $S_\mathcal{O}(x) = 0$ for $|x|>(\mathcal{O}+1)/2$;
  \item smoothness: $d^k S_\mathcal{O}(x)/dx^k$ must be defined and continuous for $\forall k \ge0$ and for $\forall x$.
\end{enumerate}

\noindent In the Appendix~\ref{sec:shapeFunctions}, we derive the expressions for shape functions of up to $\mathcal{O}\leq 5$, and in Fig.~\ref{fig:shape_functions} we show the plots for up to $\mathcal{O}\leq 11$ (which is the largest $\mathcal{O}$ implemented in \entity).

Since Eq.~\eqref{eq:chargeCont} is linear, it is enough to describe the current deposition scheme for a single particle; simple summation of individual current contributions will automatically satisfy the same equation. 
Solving the discretized version of Eq.~\eqref{eq:chargeCont} reduces to solving a linear recurrent set of equations for the discretized components of current densities:

\begin{equation}
  \label{eq:currentsSystem}
  \begin{aligned}
    J^{x}_{i+3/2,j,k} &= J^{x}_{i+1/2,j,k} - \frac{q_s \delta x}{\Delta t} \mathcal{W}^x_{i,j,k}, \\
    J^{y}_{i,j+3/2,k} &= J^{y}_{i,j+1/2,k} - \frac{q_s \delta y}{\Delta t} \mathcal{W}^y_{i,j,k}, \\
    J^{z,\alpha}_{i,j,k+3/2} &= J^{x,\alpha}_{i,j,k+1/2} - \frac{q_s \delta z}{\Delta t} \mathcal{W}^z_{i,j,k},
  \end{aligned}
\end{equation}

\noindent where we assume that the particle of charge $q_s$ is moving from point $(x_1,y_1,z_1)$ to $(x_2,y_2,z_2)\equiv(x_1+\delta x,y_1+\delta y,z_1+\delta z)$ during a timestep of duration $\Delta t$. Note here that the current density components are staggered, similar to the electric field components (edge-centered) with respect to the corner of each cell $(i, j, k)$. 
The $\mW$ coefficients are defined as follows (\citealt{Esirkepov2001}):

\begin{equation}
  \label{eq:Wcoeffs}
  \begin{aligned}
    \begin{split}
      \mW^x_{i,j,k} =& \delta \mS^{x;i} \left( \mS_1^{y;j} \mS_1^{z;k} + \frac{1}{2} \delta\mS^{y;j} \mS_1^{z;k} \right.\\
      &+ \left. \frac{1}{2} \mS_1^{y;j} \delta\mS^{z;k} + \frac{1}{3} \delta \mS^{y;j} \delta \mS^{z;k} \right),
    \end{split}
    \\
    \begin{split}
      \mW^y_{i,j,k} =& \delta\mS^{y;j} \left( \mS_1^{x;i} \mS_1^{z;k} + \frac{1}{2} \delta\mS^{x;i} \mS_1^{z;k} \right.\\
      &+ \left. \frac{1}{2} \mS_1^{x;i} \delta\mS^{z,j} + \frac{1}{3} \delta \mS^{x;i} \delta \mS^{z,j} \right),
    \end{split}
    \\
    \begin{split}
      \mW^z_{i,j,k} =& \delta\mS^{z;k} \left( \mS_1^{x;i} \mS_1^{y;j} + \frac{1}{2} \delta\mS^{x;i} \mS_1^{y;j} \right. \\
      &+ \left. \frac{1}{2} \mS_1^{x;i} \delta\mS^{y;j} + \frac{1}{3} \delta \mS^{x;i} \delta \mS^{y;j} \right),
    \end{split}
  \end{aligned}
\end{equation}

\noindent where we use the notations:

\begin{align*}
  \mS^{x;i}_1 &\equiv \mS_\mathcal{O}(x_1-x^i),\\
  \delta \mS^{x;i} &\equiv \mS_\mathcal{O}(x_2-x^i) - \mS_\mathcal{O}(x_1-x^i),\\
\end{align*}

\noindent and similarly for $y$ and $z$. Here, $x^i$, $y^j$, $z^k$ are the specific points on the grid along each individual direction. The sum $\mW_{i,j,k} \equiv \mW^{x}_{i,j,k}+\mW^{y}_{i,j,k}+\mW^{z}_{i,j,k}$, in essence, represent the time-derivative of the charge density at the grid point $(i,j,k)$:

\begin{equation}
  \label{eq:deltaCharge}
  \begin{split}
    \mW_{i,j,k} &= \mS^{\mathtt{3D}}_\mathcal{O}(\bm{x}_2-\bm{x}^{i,j,k}) - \mS^{\mathtt{3D}}_\mathcal{O}(\bm{x}_1-\bm{x}^{i,j,k}) \\
    &\equiv  \mS_\mathcal{O}(x_2-x^i)\mS_\mathcal{O}(y_2-y^j)\mS_\mathcal{O}(z_2-z^k) -\\
    & ~~~ \mS_\mathcal{O}(x_1-x^i)\mS_\mathcal{O}(y_1-y^j)\mS_\mathcal{O}(z_1-z^k).
  \end{split}
\end{equation}

Together with the coefficients defined in \eqref{eq:Wcoeffs}, the chain of recurrence equations \eqref{eq:currentsSystem} on each individual component of the current can be resolved, since the coefficients $\mW$ that are sufficiently far from the particle's position are all zero (the normalization of the shape function also ensures the solution exists and is unique for a given set of $\mW$'s).

\begin{figure}
  \centering
  \begin{tikzpicture}
  \usetikzlibrary{arrows.meta,calc}

  \def\crosssize{0.12}

  \def\scale{3}
  \def\gapH{0.2}
  \def\xInit{0.55}
  \def\yInit{0.22}
  \def\xFinal{0.25}
  \def\yFinal{0.65}

  \newcommand{\drawcross}[2]{
    \draw[#1,line width=1.2pt] ($ (#2)+(-\crosssize,-\crosssize) $) -- ($ (#2)+(\crosssize,\crosssize) $);
    \draw[#1,line width=1.2pt] ($ (#2)+(-\crosssize,\crosssize) $) -- ($ (#2)+(\crosssize,-\crosssize) $);
  }

  \coordinate (cornerBtmRect) at ({0},{0});
  \coordinate (cornerTopRect) at ({0},{\scale*(1+\gapH)});
  \coordinate (pInit) at ({\xInit*\scale},{\yInit*\scale});
  \coordinate (pFinal) at ({(1+\xFinal)*\scale},{\yFinal*\scale});
  \coordinate (pZigzagMid) at (\scale, {0.5*(\yInit+\yFinal)*\scale});
  \coordinate (pEsirkepovL) at ({\xInit*\scale},{\yFinal*\scale});
  \coordinate (pEsirkepovR) at ({(1+\xFinal)*\scale},{\yInit*\scale});

  \node[anchor=south, align=center] at ($(cornerTopRect) + ({\scale},{\scale})$) {Zigzag};

  \draw[line width=1.15pt] (cornerTopRect) rectangle ++({\scale*2},{\scale});
  \draw[line width=1.15pt] ($(cornerTopRect) + ({\scale},0)$) -- ++(0,\scale);

  \drawcross{cyan!85!black}{$(cornerTopRect)+(pInit)$}
  \drawcross{magenta!85!black}{$(cornerTopRect)+(pFinal)$}
  \drawcross{black}{$(cornerTopRect)+(pZigzagMid)$}
  \draw[-{Stealth},densely dashed,gray!70,line width=1.0pt] ($(cornerTopRect)+(pInit)$) -- ($(cornerTopRect)+(pZigzagMid)$);
  \draw[-{Stealth},densely dashed,gray!70,line width=1.0pt] ($(cornerTopRect)+(pZigzagMid)$) -- ($(cornerTopRect)+(pFinal)$);

  \node[anchor=south, align=center] at ($(cornerBtmRect) + ({\scale},{\scale})$) {Esirkepov};

  \draw[line width=1.15pt] (cornerBtmRect) rectangle ++({\scale*2},{\scale});
  \draw[line width=1.15pt] ($(cornerBtmRect) + ({\scale},0)$) -- ++(0,\scale);

  \drawcross{cyan!85!black}{$(cornerBtmRect)+(pInit)$}
  \drawcross{magenta!85!black}{$(cornerBtmRect)+(pFinal)$}
  \drawcross{black}{$(cornerBtmRect)+(pEsirkepovL)$}
  \drawcross{black}{$(cornerBtmRect)+(pEsirkepovR)$}
  \draw[-{Stealth},densely dashed,gray!40,line width=1.0pt] ($(cornerBtmRect)+(pInit)$) -- ($(cornerBtmRect)+(pEsirkepovL)$);
  \draw[-{Stealth},densely dashed,gray!40,line width=1.0pt] ($(cornerBtmRect)+(pInit)$) -- ($(cornerBtmRect)+(pEsirkepovR)$);
  \draw[-{Stealth},densely dashed,gray!40,line width=1.0pt] ($(cornerBtmRect)+(pEsirkepovL)$) -- ($(cornerBtmRect)+(pFinal)$);
  \draw[-{Stealth},densely dashed,gray!40,line width=1.0pt] ($(cornerBtmRect)+(pEsirkepovR)$) -- ($(cornerBtmRect)+(pFinal)$);

\end{tikzpicture}
  \caption{Decomposition of the implicit particle motion (from cyan cross to magenta cross) into segments in \zz and Esirkepov's algorithm.}
  \label{fig:zz-esirkepov}
\end{figure}
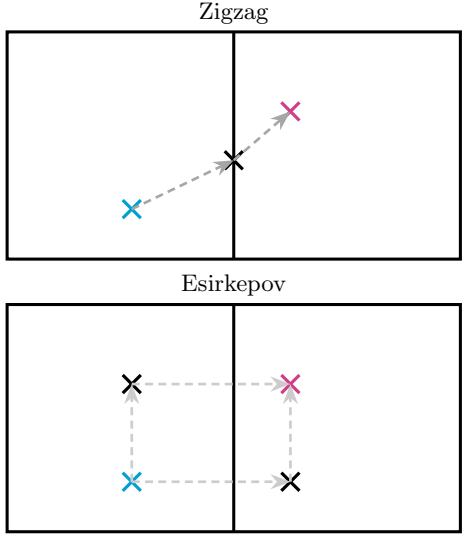

One important sidenote in this context is that the derivation of \eqref{eq:currentsSystem} relies on a specific decomposition of the motion of the particle. In other words, given the start and end points of the particle, and thus a specific value for the change in the charge density given by Eq.~\eqref{eq:deltaCharge}, one can find arbitrarily many values for the current density components that satisfy the charge continuity \eqref{eq:chargeCont}, by decomposing the motion of the particle from its initial position to the final one into multiple segments. As an example, we show the comparison between the so-called \zz algorithm described by \cite{Umeda2003} with Esirkepov's algorithm, which we employ, in Fig.~\ref{fig:zz-esirkepov}. While the sum of all $\sum_{i,j,k}\mW_{i,j,k}$ is, ultimately, zero in both \zz and Esirkepov's algorithms, individual values are different, resulting in slightly different deposited currents when comparing the 1\st-order Esirkepov deposition with the \zz algorithm.

In \entity, we implement shape functions up to 11\thu-order. We note, however, that 10\thu~and 11\thu-order require double precision operations\footnote{This can be enabled during the compilation of the code by passing the \texttt{-D precision=double} flag to \texttt{cmake}.} due to the small numerical values of the shape function at the stencil boundaries. This limits its use on certain GPU architectures, which do not support double-precision arithmetic. Therefore, we limit the majority of our testing to 1\st-to-9\thu~order to ensure the presented results are reproducible on all hardware.
We have verified the accuracy of 10\thu, and 11\thu-order independently and generally find only limited improvements compared to 9\thu-order.

\subsection{Field interpolation with high-order shape functions}

When using shape functions of order $\mathcal{O}$ for current deposition, for the scheme to be consistent, it is additionally required to use the same shape functions when interpolating the fields from the grid to the particle positions, as indicated by the Eqs.~\eqref{eq:velUpd} and~\eqref{eq:fieldInterp}.

In previous versions of \entity, we employed a node-based indirect interpolation scheme, where the electric and magnetic fields are first interpolated from edges and faces, respectively, to the nodes, and only then interpolated to the particle position. While strictly speaking, this scheme is inconsistent with the current deposition scheme even for the first-order shape functions, it has been widely used in the PIC community: see, e.g., \texttt{Tristan-MP} \citep{Buneman1993, Spitkovsky2005}, \texttt{Tristan v2} \citep{tristanv2} and \texttt{Zeltron} \citep{Cerutti2013}.

In the present version of \entity, the electromagnetic fields are instead interpolated directly from edges and faces to the particle position\footnote{For an intuitive visualisation of which components contribute to a particle, see \url{https://entity-toolkit.github.io/wiki/content/fun/particle-shapes/}.} by directly computing the convolution from Eq.~\eqref{eq:fieldInterp}:

\begin{widetext}
\centering
\resizebox{0.8\textwidth}{!}{%
  $  \begin{aligned}
      E^x_p &= \sum\limits_{
        \subalign{
          \Delta i&\in [-\lceil\mathcal{O}/2\rceil;\lceil\mathcal{O}/2\rceil] \\
          \Delta j&\in [-\lceil(\mathcal{O}+1)/2\rceil+1;\lceil(\mathcal{O}+1)/2\rceil] \\
          \Delta k&\in [-\lceil(\mathcal{O}+1)/2\rceil+1;\lceil(\mathcal{O}+1)/2\rceil]
        }
      }
      \mS(x_p - x^{i_p+\Delta i+1/2}) \mS(y_p - y^{j_p+\Delta j}) \mS(z_p - z^{k_p+\Delta k}) E^x_{i_p+\Delta i+1/2, j_p+\Delta j, k_p+\Delta k}, \\
      E^y_p &= \sum\limits_{
        \subalign{
          \Delta i&\in [-\lceil(\mathcal{O}+1)/2\rceil+1;\lceil(\mathcal{O}+1)/2\rceil] \\
          \Delta j&\in [-\lceil\mathcal{O}/2\rceil;\lceil\mathcal{O}/2\rceil] \\
          \Delta k&\in [-\lceil(\mathcal{O}+1)/2\rceil+1;\lceil(\mathcal{O}+1)/2\rceil]
        }
      }
      \mS(x_p - x^{i_p+\Delta i}) \mS(y_p - y^{j_p+\Delta j+1/2}) \mS(z_p - z^{k_p+\Delta k}) E^y_{i_p+\Delta i, j_p+\Delta j+1/2, k_p+\Delta k}, \\
      E^z_p &= \sum\limits_{
        \subalign{
          \Delta i&\in [-\lceil(\mathcal{O}+1)/2\rceil+1;\lceil(\mathcal{O}+1)/2\rceil] \\
          \Delta j&\in [-\lceil(\mathcal{O}+1)/2\rceil+1;\lceil(\mathcal{O}+1)/2\rceil] \\
          \Delta k&\in [-\lceil\mathcal{O}/2\rceil;\lceil\mathcal{O}/2\rceil]
        }
      }
      \mS(x_p - x^{i_p+\Delta i}) \mS(y_p - y^{j_p+\Delta j}) \mS(z_p - z^{k_p+\Delta k+1/2}) E^z_{i_p+\Delta i, j_p+\Delta j, k_p+\Delta k+1/2}, \\
      B^x_p &= \sum\limits_{
        \subalign{
          \Delta i&\in [-\lceil(\mathcal{O}+1)/2\rceil+1;\lceil(\mathcal{O}+1)/2\rceil] \\
          \Delta j&\in [-\lceil\mathcal{O}/2\rceil;\lceil\mathcal{O}/2\rceil] \\
          \Delta k&\in [-\lceil\mathcal{O}/2\rceil;\lceil\mathcal{O}/2\rceil]
        }
      }
      \mS(x_p - x^{i_p+\Delta i}) \mS(y_p - y^{j_p+\Delta j+1/2}) \mS(z_p - z^{k_p+\Delta k+1/2}) B^x_{i_p+\Delta i, j_p+\Delta j+1/2, k_p+\Delta k+1/2}, \\
      B^y_p &= \sum\limits_{
        \subalign{
          \Delta i&\in [-\lceil\mathcal{O}/2\rceil;\lceil\mathcal{O}/2\rceil] \\
          \Delta j&\in [-\lceil(\mathcal{O}+1)/2\rceil+1;\lceil(\mathcal{O}+1)/2\rceil] \\
          \Delta k&\in [-\lceil\mathcal{O}/2\rceil;\lceil\mathcal{O}/2\rceil]
        }
      }
      \mS(x_p - x^{i_p+\Delta i+1/2}) \mS(y_p - y^{j_p+\Delta j}) \mS(z_p - z^{k_p+\Delta k+1/2}) B^y_{i_p+\Delta i+1/2, j_p+\Delta j, k_p+\Delta k+1/2}, \\
      B^z_p &= \sum\limits_{
        \subalign{
          \Delta i&\in [-\lceil\mathcal{O}/2\rceil;\lceil\mathcal{O}/2\rceil] \\
          \Delta j&\in [-\lceil\mathcal{O}/2\rceil;\lceil\mathcal{O}/2\rceil] \\
          \Delta k&\in [-\lceil(\mathcal{O}+1)/2\rceil+1;\lceil(\mathcal{O}+1)/2\rceil]
        }
      }
      \mS(x_p - x^{i_p+\Delta i+1/2}) \mS(y_p - y^{j_p+\Delta j+1/2}) \mS(z_p - z^{k_p+\Delta k}) B^z_{i_p+\Delta i+1/2, j_p+\Delta j+1/2, k_p+\Delta k},
    \end{aligned}$
 } %
\end{widetext}

\noindent where $(i_p, j_p, k_p)$ is the index of the grid cell in which the particle is located, and $\Delta i$, $\Delta j$, $\Delta k$ are the indices of the grid points that contribute to the interpolation of the fields to the particle position. The limits of summation are defined by the order of the shape function $\mathcal{O}$, and are chosen such that they include all grid points for which $\mathcal{S}_\mathcal{O}$ is non-zero.

\subsection{Field solver with optimized stencils}
\label{sec:methods_fieldsolver}

Faraday's and Amp\`ere's equations,~\eqref{eq:faraday} and \eqref{eq:ampere}, are solved by discretizing the spatial derivatives with finite difference stencils, and the time derivative with a Leapfrog scheme. The most common technique to solve these equations was introduced by \citet{Yee1966} and employs a staggered grid between $\bm{E}$- and $\bm{B}$-fields --- the Yee-mesh --- constructing a second-order accurate central finite difference stencil between two adjacent grid-points:

\begin{equation*}
  \begin{aligned}
    \partial_t \bm{X} & \rightarrow
    D_t\left( \bm{X}_{i,j,k}^{n} \right) =
    \frac{1}{\Delta t} \left( \bm{X}_{i,j,k}^{n+\frac{1}{2}} - \bm{X}_{i,j,k}^{n-\frac{1}{2}} \right), \\
    \nabla_x & \rightarrow
    D_x\left( \bm{X}_{i,j,k}^{n} \right) =
    \frac{1}{\Delta x} \left( \bm{X}_{i+\frac{1}{2},j,k}^{n} - \bm{X}_{i-\frac{1}{2},j,k}^{n} \right),
  \end{aligned}
\end{equation*}

\noindent where $\bm{X}$ is either $\bm{E}$ or $\bm{B}$, and the indices $i,j,k$ and $n$ correspond to the spatial and temporal grid points (staggered with respect to cell corners), respectively. We further refer to this scheme as the \texttt{Yee} stencil.

Depending on the problem considered, one may seek improved accuracy or enhanced stability by modifying the numerical dispersion properties of the scheme \citep[e.g.][]{Vay2011}. One way to improve upon this approach is to include additional adjacent grid points, thereby widening the stencil. This naturally leads to a broad class of higher-order finite-difference schemes, each offering distinct advantages depending on the physical regime of interest. Some approaches rely on customized higher-order stencils for both of the Eqs.~\eqref{eq:ampere} and \eqref{eq:faraday}, at the expense of requiring nonlocal charge-conservation procedures, e.g.~\citet{Li2017}. Others apply higher-order discretizations only to Faraday's equation, \eqref{eq:faraday}, thereby preserving the standard current deposition scheme, e.g.~\cite{Cole1997, Cole2002, Karkkainen2006, Lehe2013}. For computational efficiency and broader generality, we implement the latter generalized stencil in \entity~and discuss here the properties of a large variety of stencils optimized for various regimes of interest following \citet{Blinne2018}.

The field stencil for the right-hand side of Faraday's equation can be generalized to
\begin{align}
  \begin{split}
    D_x^*(\bm{X}_{i,j,k}^n) =& \frac{\alpha_x}{\Delta x} \left( \bm{X}_{i+\frac{1}{2},j,k}^n - \bm{X}_{i-\frac{1}{2},j,k}^n \right) \\
    +& \frac{\delta_x}{\Delta x} \left( \bm{X}_{i+\frac{3}{2},j,k}^n - \bm{X}_{i-\frac{3}{2},j,k}^n \right) \\
    +& \frac{\beta_{xy}}{\Delta x} \left( \bm{X}_{i+\frac{1}{2},j+1,k}^n - \bm{X}_{i-\frac{1}{2},j+1,k}^n \right)\\
    +& \frac{\beta_{xy}}{\Delta x} \left( \bm{X}_{i+\frac{1}{2},j-1,k}^n - \bm{X}_{i-\frac{1}{2},j-1,k}^n \right)\\
    +& \frac{\beta_{xz}}{\Delta x} \left( \bm{X}_{i+\frac{1}{2},j,k+1}^n - \bm{X}_{i-\frac{1}{2},j,k+1} \right)\\
    +& \frac{\beta_{xy}}{\Delta x} \left( \bm{X}_{i+\frac{1}{2},j,k-1}^n - \bm{X}_{i-\frac{1}{2},j,k-1}^n \right),
  \end{split}
\end{align}
and similar for $D_y^*$, and $D_z^*$, which we omit in the interest of brevity.
The $\alpha$ coefficients are defined as
\begin{equation}
  \label{eq:alpha_xyz}
  \begin{aligned}
    \alpha_x &= 1 - 2\beta_{xy} - 2\beta_{xz} - 3\delta_x,\\
    \alpha_y &= 1 - 2\beta_{yx} - 2\beta_{yz} - 3\delta_y,\\
    \alpha_z &= 1 - 2\beta_{zx} - 2\beta_{zy} - 3\delta_z.
  \end{aligned}
\end{equation}
It follows trivially that if $\delta_i = 0$ and $\beta_{ij} = 0$, the field stencil falls back to the standard \texttt{Yee} stencil.

\begin{table*}[]
  \centering
  \caption{Parameters for the 2D field stencils as provided in \citet{Blinne2018}. 3D field stencils are constructed using the optimisation process described in Sec.~\ref{sec:methods_fieldsolver}. Stencils marked with $\leftrightarrow$ are optimized for a specific direction. The CFL definition is consistent with the reference publications. For usage in \entity~note that $\mathrm{CFL}_\entity = \mathrm{CFL} \times \sqrt{N_\mathrm{dim}}$, where $N_\mathrm{dim}$ is the dimensionality of the simulation.}
  \begin{tabular}{l|cccccccccc}
    \toprule
    Solver & CFL & $\delta_x$ & $\delta_y$ & $\delta_z$ & $\beta_{xy}$ & $\beta_{yx}$ & $\beta_{xz}$ & $\beta_{zx}$ & $\beta_{yz}$ & $\beta_{zy}$\\
    \midrule
    \textbf{2D}&&&&&&&&\\
    \texttt{Yee} & -- & 0 & 0 & 0 & 0 & 0 & 0 & 0 & 0 & 0 \\
    \texttt{Cowan} & 0.999 & 0 & 0 & 0 & 0.125 & 0.125 & 0 & 0 & 0 & 0 \\
    \texttt{Lehe}$\leftrightarrow$ & 0.96 & -0.021 & 0 & 0 & 0.125 & 0.125 & 0 & 0 & 0 & 0\\
    \texttt{min1} & 0.97/$\sqrt{2}$ & -0.125 & -0.125 & 0 & 0.11 & 0.11 & 0 & 0 & 0 & 0\\
    \texttt{min2} & 0.95/$\sqrt{2}$ & -0.013 & -0.013 & 0 & -0.013 & -0.013 & 0 & 0 & 0 & 0\\
    \texttt{min3} & 0.5 & -0.065 & -0.065 & 0 & -0.065 & -0.065 & 0 & 0 & 0 & 0\\
    \texttt{min4} & 0.1 & -0.125 & -0.125 & 0 & -0.125 & -0.125 & 0 & 0 & 0 & 0\\
    \texttt{min5} & 0.96 & -0.017 & -0.017 & 0 & 0.133 & 0.133 & 0 & 0 & 0 & 0\\
    \texttt{min6}$\leftrightarrow$& 0.999 & -0.0005 & 0 & 0 & 0.128 & 0.128 & 0 & 0 & 0 & 0\\
    \midrule
    \textbf{3D}&&&&&&&&\\
    \texttt{Yee} & -- & 0 & 0 & 0 & 0 & 0 & 0 & 0 & 0 & 0 \\
    \texttt{Lehe3D}$\leftrightarrow$ & 0.96 & -0.021 & 0 & 0 & 0.125 & 0.125 & 0.125 & 0.125 & 0.125 & 0.125\\
    \texttt{min3D1} & 0.5 & -0.0273 & -0.0273 & -0.0273 & -0.0273 & -0.0273 & -0.0273 & -0.0273 & -0.0273 & -0.0273 \\
    \texttt{min3D2} & 0.5 & -0.1704 & -0.1704 & -0.1704 & 0.05801 & 0.05801 & 0.05801 & 0.05801 & 0.05801 & 0.05801 \\
    \texttt{min3D3} & $1/\sqrt{3}$ & -0.16 & -0.16 & -0.16 & 0.08 & 0.08 & 0.08 & 0.08 & 0.08 & 0.08 \\
    \bottomrule
  \end{tabular}
  \label{tab:blinne_coefficients}
\end{table*}

This generalized stencil can be used to reduce the numerical dispersion by optimizing the stencil for a given dispersion relation.
Numerical dispersion is introduced by discretizing Maxwell's equations on a grid, and can lead to numerical light waves traveling faster or slower on the discretized grid than the speed of light in vacuum.
Once relativistic particles are introduced to this grid, they may travel faster than the numerical speed of light on said grid and with that experience numerical Cherenkov effects \citep[e.g.][]{Boris1973, Godfrey1974, Cormier-Michel2008, Vay2011, Blinne2018, Lu2020, Filipovic2022, Huddleston2025}.

One way to mitigate this is by choosing the coefficients given in Eqs.~\eqref{eq:alpha_xyz} to minimize the error between the physical and numerical speed of light, by tuning the dispersion relation of the grid.
The dispersion relation of the grid is given as
\begin{equation}
  s_\omega^2 = s_x^2 A_x + s_y^2 A_y + s_z^2 A_z
  \label{eq:dispersion}
\end{equation}
with
\begin{align}
  s_\omega &= \frac{\sin\left(\frac{1}{2} \omega \Delta t \right)}{c \Delta t} \\
  s_{\{x,y,z\}} &= \frac{\sin\left( \frac{1}{2} k_{\{x,y,z\}} \Delta {\{x,y,z\}} \right)}{\Delta {\{x,y,z\}}}
\end{align}
\noindent and
\begin{align}
  \begin{split}
    A_x &= \alpha_x + 2 \beta_{xy} \cos(k_y \Delta y) + 2 \beta_{xz} \cos(k_z \Delta z) \\
    &+ \delta_x(1 + 2 \cos(k_x \Delta x)), \\
    A_y &= \alpha_y + 2 \beta_{yx} \cos(k_x \Delta x) + 2 \beta_{yz} \cos(k_z \Delta z) \\
    &+ \delta_y(1 + 2 \cos(k_y \Delta y)), \\
    A_z &= \alpha_z + 2 \beta_{zx} \cos(k_x \Delta x) + 2 \beta_{zy} \cos(k_y \Delta y) \\
    &+ \delta_z(1 + 2 \cos(k_z \Delta z)).
  \end{split}
\end{align}

The dispersion relation in Eq.~\eqref{eq:dispersion} remains well defined if
\begin{equation}
  0 \leq \left(c \Delta t\right)^2 \left( s_x^2 A_x + s_y^2 A_y + s_z^2 A_z \right) \leq 1
\end{equation}
for $0 \leq k_i \leq \pi$.
Moving $\left(c \Delta t\right)^2$ to the right inequality lets us identify it as the standard Courant-Friedrichs-Lewy (CFL) condition, which is an integral component of the optimisation procedure \citep[for a discussion see][]{Cowan2013}.

Following \citet{Blinne2018}, the optimal field stencils are obtained by minimizing the weighted deviation of the numerical phase velocity from the physical speed of light in Fourier space:

\begin{align}
  f_w(\omega) = \int_0^\pi d\tilde{k}_x \int_0^\pi d\tilde{k}_y \int_0^\pi d\tilde{k}_z \,w(\tilde{k}) \, \left( \frac{\tilde{\omega}}{c\tilde{k}} - 1\right)^2,
\end{align}

\noindent where $\tilde{k}_x = k_x \Delta x$, $\tilde{k}_y = k_y \Delta y$, $\tilde{k}z = k_z \Delta z$, $\tilde{k}^2 = \tilde{k}_x^2 + \tilde{k}_y^2 + \tilde{k}_z^2$, and $\tilde{\omega} = \omega \Delta t$. The weight function $w(\tilde{k})$ is problem-specific: for isotropic stencils such as \texttt{min1}–\texttt{min4}, \citet{Blinne2018} use $w(\tilde{k}) = \tilde{k}^2$, comparable to results obtained for a uniform weighting $w(\tilde{k}) = 1$. To reduce dispersion along a specific axis, the weight can be localized as $w(\tilde{k}) = \tilde{k}^2 \exp\left(- k_\perp^2 / \sigma^2 \right)$; for example, $\sigma = 0.1$ is used for the \texttt{min6} stencil.

Additionally, the generalized stencil automatically enforces $\bm{D}^\star \cdot \bm{B}\,=\,0$ throughout the simulation, provided that it is satisfied at the initial timestep \citep[in the absence of explicit divergence cleaning, e.g.][]{Marder1987, Langdon1992, Munz2000, Pfeiffer2015}. However, enforcing $\bm{D}^\star \cdot \bm{B}\,=\,0$ initially is generally more subtle than imposing $\bm{D} \cdot \bm{B}\,=\,0$. For such configurations, one can show that the following equivalence holds: 
\begin{equation} \left( \bm{D} \cdot \bm{B}\,=\,0 \Rightarrow \bm{D}^\star \cdot \bm{B}\,=\,0 \right) \iff \quad \beta_{ij} = \delta_j \label{eq:stencil_divB} 
\end{equation} 
for $i,j = x,y,z$ \cite[see discussion in Sec. 2.2 of][]{Blinne2018}. 

For convenience, we summarize the optimized 2D stencils listed in \citet{Blinne2018} in Table~\ref{tab:blinne_coefficients}.
We omit \texttt{NDFX} \citep{Pukhov1999} in favor of \texttt{Cowan} \citep{Cowan2013}, since it only differs in \texttt{CFL}.
The \texttt{Lehe} stencil is taken from \citet{Lehe2013}, \texttt{min1}-\texttt{min6} are introduced in \citet{Blinne2018}\footnote{They provide a public version of their optimization tool at \url{https://github.com/Ablinne/optimize-stencil}}.

We introduce the 3D stencils \texttt{min3D1}, \texttt{min3D2}, and \texttt{min3D3}, constructed with the optimisation method discussed above. The \texttt{Lehe3D} stencil is a simple 3D extension of the standard \texttt{Lehe} stencil, and with that is optimized towards a specific direction.

Stencils satisfying Eq.~\eqref{eq:stencil_divB} are \texttt{Yee}, \texttt{min2}, \texttt{min3}, \texttt{min4}, and \texttt{min3D1}. To use the stencils in \entity~one needs to multiply the CFL value by $\sqrt{N_\mathrm{dim}}$ to adhere to \entity's timestep definition.

Finally, it is worth stressing again for clarity that the values of \texttt{CFL}, $\delta_i$, and $\beta_{ij}$ are linked by construction. It is therefore not advisable to change the \texttt{CFL} for a given stencil. Instead, a new stencil should be constructed if a specific \texttt{CFL} is required.

\section{Convergence tests}
\label{sec:convergence}

Both the high-order shape functions and the advanced Faraday stencils have been shown to considerably reduce numerical noise and increase convergence in test cases. We now present a number of tests to verify the improvements compared to the previous behavior of \entity.

\subsection{High-order shape functions}
We first focus on convergence improvement when using high-order shape functions.

\subsubsection{Charge conservation}

\begin{figure}
  \centering
  \includegraphics[width=\linewidth]{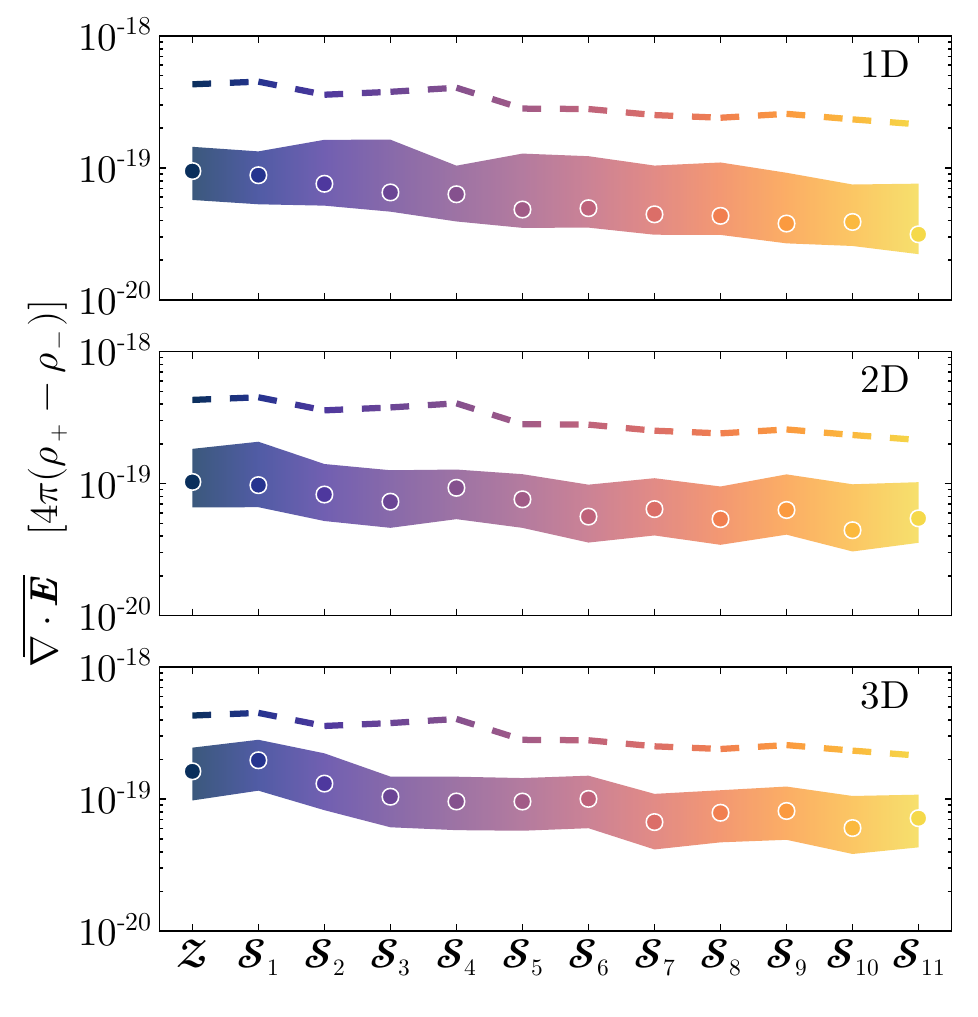}
  \caption{Charge-conservation of an initially neutral plasma at rest. We show the mean value of the total charge density (dots) and its standard deviation (band) over an evolution of $t = 10^5 \: \omega_p^{-1}$. The dashed line indicates the maximum over the course of the simulation.}
  \label{fig:divE_error}
\end{figure}

First, we want to ensure that our updated current deposition scheme is indeed charge-conserving. For this, we initialize a domain filled with a uniform pair-plasma with a temperature $\theta = 10^{-3} \: m_e c^2$, where $m_e$ is the electron mass. We initialize the plasma at rest in 1D, 2D, and 3D, using $65536$, $256^2$, and $32^3$ cells, respectively. We resolve the system with $25.6$ cells per skin-depth and employ periodic boundary conditions. The simulation is evolved for $t = 10^5 \: \omega_p^{-1}$, with $\omega_p = \sqrt{\frac{4\pi n_0 q^2}{m_e}}$ being the (electron) plasma frequency. We compute the volume-averaged charge, $\overline{\nabla \cdot \bm{E}}\equiv V^{-1}\int \nabla \cdot \bm{E}dV$ in intervals of $10^2 \: \omega_p^{-1}$ and show the mean and the standard deviation over the runtime in Fig.~\ref{fig:divE_error} as a function of the shape order, $\mathcal{O}$. We normalize the computed charge density with the absolute value of the total charge of both species $4\pi(|\rho_+|+|\rho_-|)$.
The dashed line indicates the maximum of $\overline{\nabla \cdot \bm{E}}$ over the course of the simulation.
Here, and in the remainder of the paper, $\mathcal{Z}$ refers to the \zz current deposit scheme, while $\mathcal{S}_1 - \mathcal{S}_{11}$ refer to the \es deposit scheme, using 1\st-11\thu-order shape functions.
This test was performed in \texttt{double precision} to include 10\thu- and 11\thu-order.

Generally, we find excellent charge conservation of the order of the truncation error.
There is a weak dependence on the shape function order, with higher orders leading to marginally improved charge conservation.

\subsubsection{Stopping power\label{sec:stopping_power}}

\begin{figure*}
  \centering
  \includegraphics[width=\textwidth]{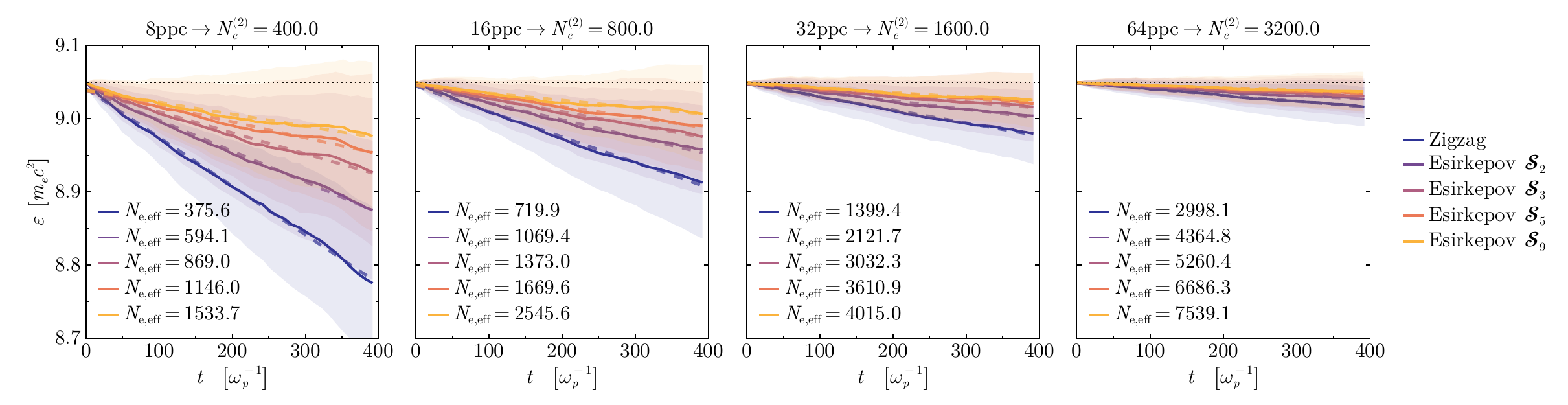}
  \caption{Energy loss rate for an electron population traveling through a neutral background plasma in 2D. Each panel refers to runs with increasing electrons in a skin-depth volume. Colors show the results for the respective shape functions. Solid lines indicate the simulation results, while dashed lines indicate a fit to the data. The labels show the effective electrons in a skin-depth volume, according to Eq.~\ref{eq:N_e}. The colored bands show the standard deviation of the energy loss rate of the relativistic electron population.}
  \label{fig:stopping_power}
\end{figure*}

Akin to the gravitational dynamic drag experienced by massive bodies moving through a background medium, high-energy particles propagating in a bath of quasi-neutral plasma experience an electrostatic stopping force. As discussed by \citet{Kato2013}, this effect is strongly enhanced for PIC macroparticles, which inevitably undersample the otherwise continuous distribution function. In a perfectly collisionless plasma under realistic astrophysical conditions, the energy loss is negligible due to the large number of real particles per skin-depth volume (about $10^{22}$ particles in an electron–proton plasma within one skin-depth volume for a density of $1\,{\rm cm}^{-3}$); however, the finite resolution of the simulation introduces an effective stopping power. This is particularly relevant in low Mach number shocks, where particle acceleration is slow \citep[e.g.][]{Guo2014a, Ha2021, Kobzar2021}. Numerically increased energy losses can stall the particle acceleration and lead to incorrect results.

\citet{Kato2013} introduced a test for the stopping power of a plasma on a discretized computational grid, which we aim to reproduce in this section. 

We initialize background plasma at rest, with a temperature of $T = 16$ keV and a mass ratio of $m_i/m_e = 1836$.
This choice differs from the $m_i/m_e = 20$ mass ratio in \citet{Kato2013}; they note that their derivation is independent of mass ratio, as long as  $m_i \gg m_e$ is fulfilled.
We find generally similar behaviour; however, in our tests, the variance between different runs decreases with larger mass ratio, hence we choose a realistic mass ratio to test the extreme limit.

We then set up 100 electron-positron pairs on top of the background plasma.
The pairs are initialized at the same position to avoid phantom charges, but only the electrons are given a constant drift velocity of $\gamma = 10$ in the positive x-direction through the background plasma.
To ensure that the stopping power is not increased by numerical Cherenkov effects, we additionally employ the \texttt{min3} field stencil (see Sec.~\ref{sec:convergence_fielstencil} for details), as it provides the best match between numerical and physical phase velocity for a drift along the x-direction \citep[see fig. 1 in][]{Blinne2018}.
We set up a 2D domain with periodic boundary conditions and a transverse size of $L = 102.4 \: d_e$ at a spatial resolution of $\Delta x = 0.1 \: d_e$.

The energy loss rate of a particle moving in the plasma can be related to the simulation resolution, in this case, the number of electrons in a skin-depth volume, following equation 34 in \citet{Kato2013}.
In 2D, we can calculate an effective number of electrons in a skin-depth volume from the measured electron energy loss rate  by solving this equation for $N_e^{(2)}$:
\begin{equation}
  N_e^{(2)} = - \frac{q_0^2}{4} \left( \frac{d\varepsilon}{dt} \right)^{-1} \: .
  \label{eq:N_e}
\end{equation}
The results of our test simulations are shown in Fig.~\ref{fig:stopping_power}.
Solid lines indicate the average energy of the relativistic electron population, as a function of time.
The surrounding bands indicate the standard deviation of the energy.
Colors correspond to the order of the shape function used in the run, with \zz being equivalent to $\mathcal{S}_1$, compared to four runs with the \es deposit, using $\mathcal{S}_2$, $\mathcal{S}_3$, $\mathcal{S}_5$, and $\mathcal{S}_9$.
Dashed lines indicate the fit to the energy loss rate, used to obtain the effective number of electrons per skin-depth volume $N_{e,\mathrm{eff}}$.
Between the panels, we incrementally increase the number of particles per cell (ppc) by a factor of two to compare different numerical resolutions.
All simulations were carried out without current filtering, as in \citet{Kato2013}.

We find that we obtain reasonable agreement between the analytic and observed energy loss in the \zz deposit.
For 8ppc, the analytic value for $N_e^{(2)}$ is recovered almost exactly, while for higher ppc the error remains of the order 10\%.

Moving to 2\nd~order \es deposit reduces the stopping power considerably, leading to a result that behaves like an effective resolution increase by almost 50\%.
This trend continues at a decreasing rate from 3\rd~to 9\thu-order, where 3\rd-order provides an effective resolution increase between a factor of one and a half and two, 5\thu-order by a factor of two to three, and 9\thu-order leads to an effective resolution increase by a factor of three to four.

While we independently verified the findings of \citet{Kato2013} in 1D, that the energy loss rate is almost independent of the shape function of the particles, extending the tests to 2D shows an impact of the shape function order on the stopping power of a plasma.
In our tests, this can be attributed to a contribution to the stopping power by numerical Cherenkov effects. We have verified that this is the case by comparing the presented results with simulations using the \texttt{Yee} and \texttt{min1} stencils, as they provide a higher and lower phase velocity, respectively, compared to the physical phase velocity for a drift along the x-direction. With the \texttt{Yee} stencil, the stopping power is increased, equivalent to a factor of two in resolution decrease. For the \texttt{min1} stencil using the \zz deposit, the stopping power is decreased equivalent to an increase in resolution by 50\%, while for the \es deposit, the stopping power is decreased equivalent to a factor of three in resolution increase. The increase is independent of the shape function order, however, only beyond 2\nd-order, when using the \texttt{min1} stencil.

\subsubsection{Numerical heating\label{sec:heating}}

\begin{figure*}
  \centering
  \includegraphics[width=\textwidth]{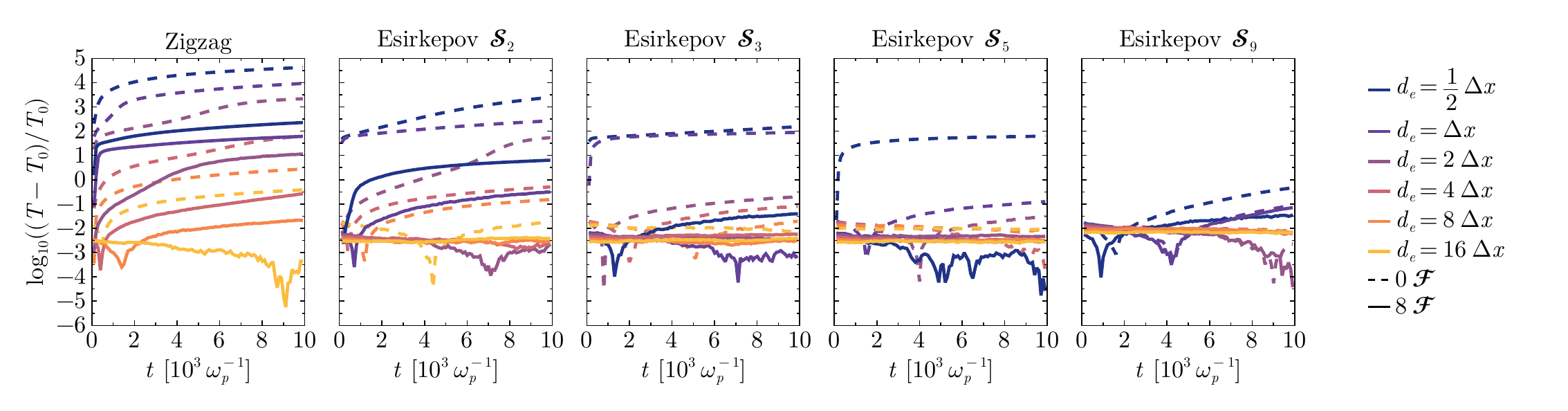}
  \caption{Numerical heating due to unresolved Debye length. The panels show the relative temperature increase over time. Each column shows the runs performed with the respective shape function order. Colors indicate the different resolution levels. Dashed lines indicate runs without current filtering, while solid lines refer to runs with 8 current filter passes.}
  \label{fig:numerical_heating}
\end{figure*}

  \begin{figure*}
    \centering
    \includegraphics[width=\textwidth]{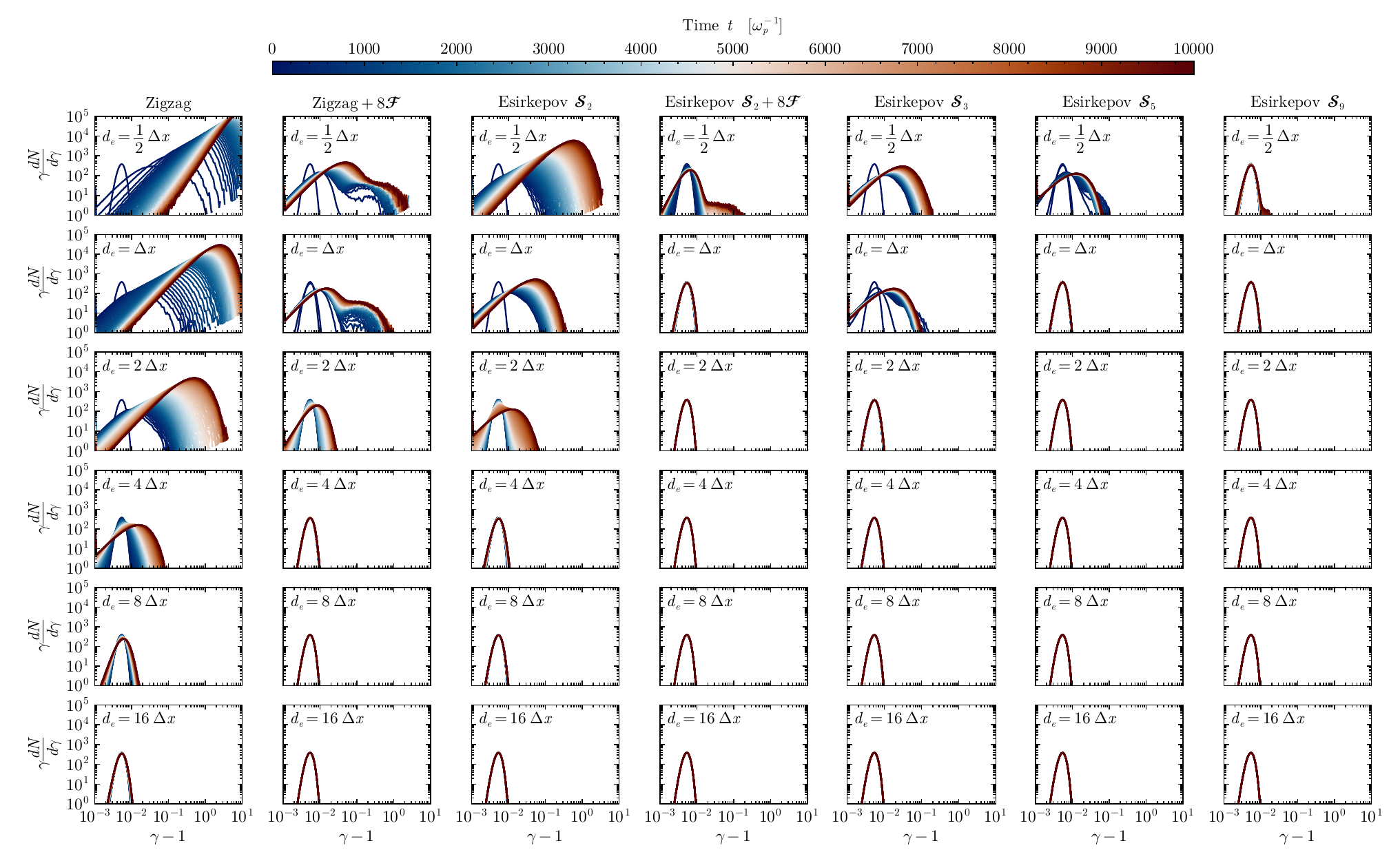}
    \caption{Spectra of particle distributions in the numerical heating test. The plot titles refer to the deposit scheme used in the simulation and whether it employs current filtering. From top to bottom, the panels show increasing grid resolution by a factor of two. The line colors indicate the time at which the calculated spectrum was measured.}
    \label{fig:numerical_heating_spectra}
  \end{figure*}

Another way to test numerical stability is to check for numerical heating.
For this we set up a cold electron-ion plasma with $T_e = T_i = 10^{-4} \: m_e c^2$ sampled with 16 particles per cell (8 per species),  $m_i = 100 \: m_e$, and a drift velocity of $v_\mathrm{drift} = 0.1 \: c$ in periodic boundary conditions. We then vary the electron skin-depth scale, $d_e$, between $1/2$ and $16$ grid cells, $\Delta x$.

For the initialized plasma to be in a stable configuration, the numerical discretization must resolve the square root of the Debye length, $\lambda_D\approx d_e\sqrt{T_e/m_e c^2} \sqrt{m_i/m_e}$, as discussed in \citet{Shalaby2017, Adams2025}. We can therefore define a stability criterion as

\begin{equation}
  T_e/m_e c^2 \gtrsim \Theta_C \equiv (\Delta x/d_e)^2,
  \label{eq:debye_stability}
\end{equation}

\noindent where $\Theta_C$ is the minimum temperature satisfying the stability criterion. If this condition is not met, i.e., $\lambda_D\lesssim \Delta x$, the plasma will heat up to increase its Debye length \citep[e.g][]{Birdsall1980}.

For our choice of parameters, Eq.~\eqref{eq:debye_stability} is satisfied when $d_e\geq 10\Delta x$. The results of this test are shown in Fig.~\ref{fig:numerical_heating}, where we show the logarithm of the relative temperature increase as a function of time. From left to right, the columns show the \zz, and the \es deposit schemes with 2\nd, 3\rd, 5\thu, and 9\thu-order shape functions. The different colors indicate the resolution levels with $d_e\in [0.5; 16] \: \Delta x$. Dashed lines show runs without additional current filtering, while solid lines show runs with 8 filter passes.

When using the \zz deposit, without current filters, we find that the criterion~\eqref{eq:debye_stability} is strictly required to ensure stability; only the test with $d_e=16\:\Delta x$ produces reasonable results with numerical heating lower than an order of unity. With the addition of current filtering, the numerical heating decreases considerably, equivalent to increasing the resolution by a factor of four.

When switching to the 2\nd-order \es deposit, we find the same behaviour already for the runs without filters, thus suggesting that the 2\nd~order \es deposit behaves like \zz deposit at roughly $\times4$ the resolution. When including filters in the 2\nd-order \es deposit, we again gain a factor of $\times4$ in effective resolution compared to 2\nd-order without filtering.

For the 3\rd-order \es deposit, we find stability behaviour equivalent to a factor of $\times 8$ in effective resolution increase. When including filters in 3\rd~order, we gain another effective resolution increase of a factor of $\times4$ compared to the run without current filtering.

In 5\thu~order, we gain an effective resolution increase of a factor of $\times 16$, while in 9\thu~order, even the plasma with an underresolved skin-depth ($d_e = 0.5\Delta x$) stays numerically stable, equivalent to an effective resolution increase of a factor of $\times 32$ compared to \zz without current filtering. When including current filtering in 5\thu~and 9\thu~order, all the tested resolution levels stay numerically stable.

To further illustrate this convergence behaviour, we show the energy distributions for all the runs in Fig.~\ref{fig:numerical_heating_spectra}. Here, rows show different resolution levels, while columns show the deposit scheme and shape order, with 8 current filter passes where indicated. When examining these spectra, we generally find the same trend as with heating alone, namely that the transition from 1\st~to 2\nd~order deposit improves the stability, similar to increasing the spatial resolution by a factor $\times 4$. Equally, including current filtering increases the stability, equivalent to another factor of $\times 4$ in spatial resolution increase. Notably, when including current filtering, which suppresses the heating for the bulk of the plasma, we observe a high-energy tail developing over time in cases where the skin-depth itself is under-resolved: $d_e\lesssim \Delta x$ (noticeable both for the \zz and the \es with lower-order shape functions).

\subsubsection{Growth of oblique modes}

\begin{figure}
  \centering
  \includegraphics[width=\linewidth]{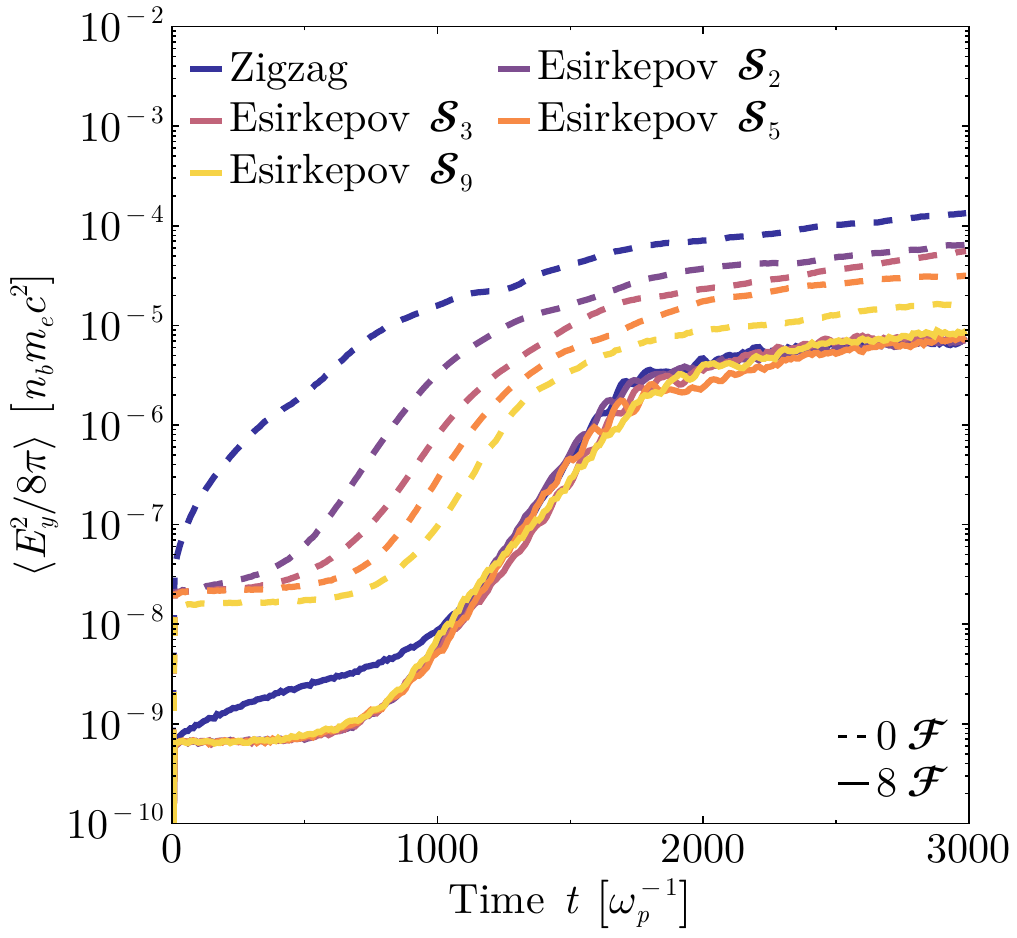}
  \caption{Growth of the $E_y$ component due to a diffuse CR beam propagating through a neutral background plasma. Different colors indicate runs with the different current deposit schemes with various shape function orders. Dashed lines indicate runs without additional current filtering, while solid lines indicate runs with 8 current filter passes.}
  \label{fig:oblique_mode_Ey}
\end{figure}

\begin{figure*}
  \centering
  \includegraphics[width=0.8\textwidth]{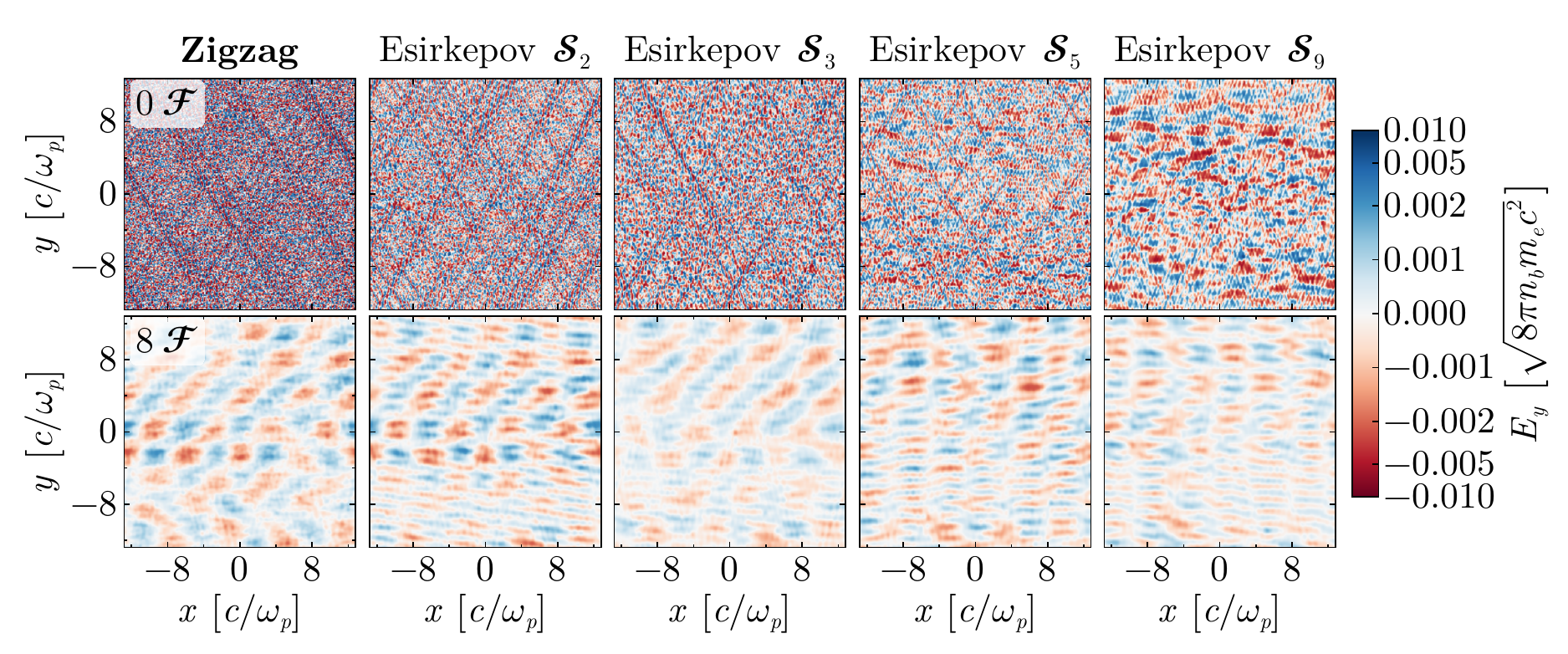}
  \caption{$E_y$ field component at $t = 1500 \: \omega_e^{-1}$ of the high obliquity modes test with a dilute CR beam propagating through a background plasma. From left to right, we show runs with the deposit scheme and shape function order indicated in the column titles. The top row shows the runs without current filtering, while the bottom row shows runs with 8 current filter passes. }
  \label{fig:oblique_mode_panels}
\end{figure*}

To further study the impact of high-order shape functions on the numerical stability of PIC, we set up a low-density quasi-neutral beam with $\gamma = 100$ propagating in a background plasma at rest in 2D. The density ratio of the beam to that of the background is $n_\mathrm{b}/n_\mathrm{bg} = 10^{-4}$. In this simulation, all particles have the same weight, and with only 16 particles per cell (ppc) in total, only 1 in 100 cells contains a particle from the beam, allowing us to study the stability of the system in a significantly under-sampled limit. We expect a numerical aliasing noise to grow in the $E_y$ and $E_z$ components of the electric field. In the interest of brevity, we will only discuss the $E_y$ component.

The growth of $E_y$ as a function of time is shown in Fig.~\ref{fig:oblique_mode_Ey}, where we plot the volume-averaged $E_y^2/8\pi$ as a fraction of the average beam rest-mass energy. The line colors indicate runs with different deposit schemes and shape function orders. Dashed lines show runs without current filters, while solid lines show runs with the addition of eight current filter passes.

Without current filtering, we see a clear suppression of numerical noise with increased shape function order. Most notably, with increased order, the initial stable configuration persists for a longer time before the onset of the numerical instability. Between \zz, and \es 1\st/9\thu~orders, we observe a decrease in the saturated energy by almost an order of magnitude. We note, however, that for this test, 6\thu-order can be considered converged, as all of the higher orders lie in a band of order unity around the 6\thu-order run, at the end of the simulation.

When including 8 current filter passes ($8 \mathcal{F}$), the overall noise level at the beginning of the simulation is reduced by more than one order of magnitude. The \zz and \es 1\st-order runs show an instant growth of the $E_y$-component, while 2\nd-to-9\thu~orders retain the initial stable configuration until $t\sim 750 \: \omega_p^{-1}$, comparable to the 9\thu~order simulation without filtering. With filtering, 2\nd~order \es can be considered converged for this test, as there is no significant difference between the higher-order runs. In fact, even the \zz and the 1\st~order \es can be considered converged, as these runs saturate at the same noise level and only differ in the initial growth phase.

This is equally evident in Fig.~\ref{fig:oblique_mode_panels}, where we show the 2D plot of $E_y$ at $t = 1500 \: \omega_p^{-1}$ for each simulation. From left to right, columns show the runs with the indicated deposit scheme and the shape function order. The top row shows the runs without current filtering, the bottom row with current filtering. While the runs with current filtering show clear signs of an oblique mode in $E_y$, the runs without current filtering are dominated by noise introduced by the numerical Cherenkov instability. This noise decreases significantly with higher-order shape functions, similar to our findings of the stopping power in Sec.~\ref{sec:stopping_power}.

\subsubsection{Bell instability}

\begin{figure}
  \centering
  \includegraphics[width=\linewidth]{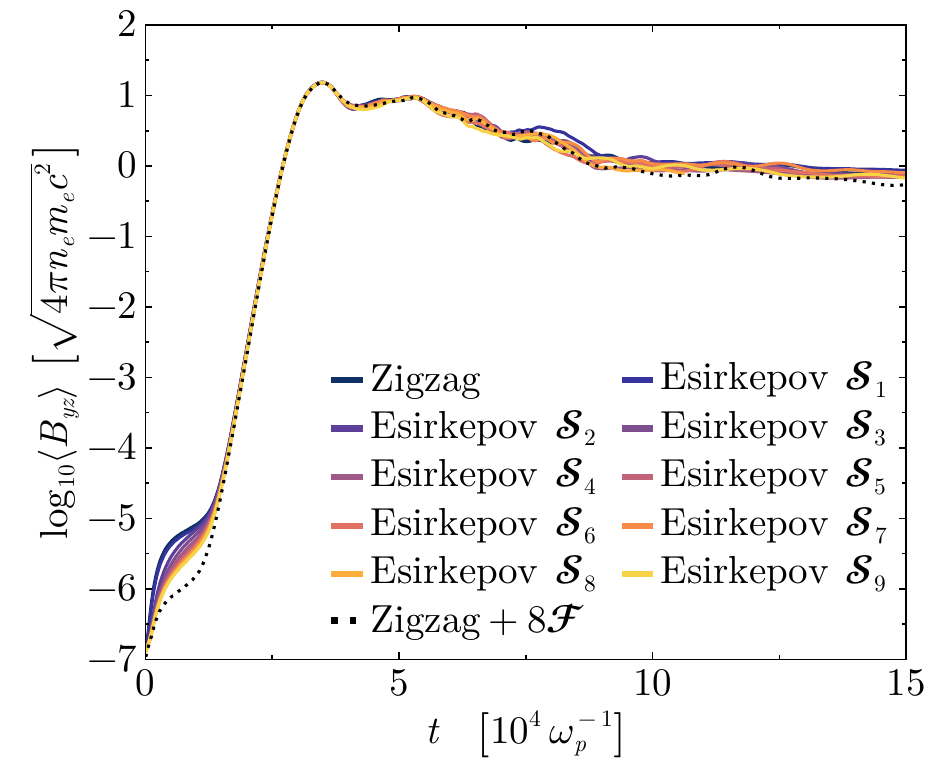}
  \caption{Perpendicular magnetic field growth as a function of time in the Bell instability test. Colors correspond to the indicated deposit schemes and interpolation orders. The dashed line shows a reference run using the \zz deposit with an additional eight current filter passes.}
  \label{fig:bell}
\end{figure}

In order to test the impact of high-order shape functions on a physically more meaningful system, we performed a series of 1D simulations of the Bell instability \citep[][]{Bell2004}. In this test, we set up a neutral electron-ion background plasma with $T_e=T_i = 10^{-4}m_e c^2$ and a mass ratio $m_i/m_e = 50$. Into this background we inject a beam of ions with the same initial temperature, drifting in $+\hat{\bm{x}}$ direction with a bulk $\gamma = 99.99625$ and a density ratio of $n_\mathrm{CR} = 10^{-2} n_\mathrm{bg}$. The densities of the ions and the electrons in both the beam and the background are chosen in such a way that the total charge density is zero. The background moves with a small velocity opposite the motion of the beam to compensate for the beam's induced electric current. The plasma is initially magnetized with $\sigma = B_x^2/(4\pi n_\mathrm{bg} c^2) = 1$, with the magnetic field pointing in the $+\hat{\bm{x}}$ direction.

We show the growth of the perpendicular magnetic field $B_{yz}$ in Fig.~\ref{fig:bell}. Since this setup is initially only magnetized along $x$, the instability grows from thermal noise along the $y\text{-}z$ direction. High-order shape functions mainly impact the turnover behavior from the initial noise-dominated growth to the actual instability. Comparing the runs, we observe a suppression of the noise-dominated growth with increasing shape function order, and a more gradual turnover to the instability-dominated growth of the magnetic field strength. However, once the instability-dominated growth phase is reached, all shape function orders quickly converge to the same growth rate, peak at the same level, and show similar saturation behavior.

When comparing to the run including eight filter passes, we see similar behavior; the magnetic field initially grows due to noise and then transitions to the instability. The filtering additionally suppresses the initial noise-dominated phase, transitioning to the instability at a lower magnetic field level. We do not see a quantifiable difference when moving to higher-order shape functions while using eight filter passes, hence we omit it here.

\subsubsection{Decaying turbulence}

\begin{figure*}[htb]
  \centering
  \includegraphics[width=\linewidth]{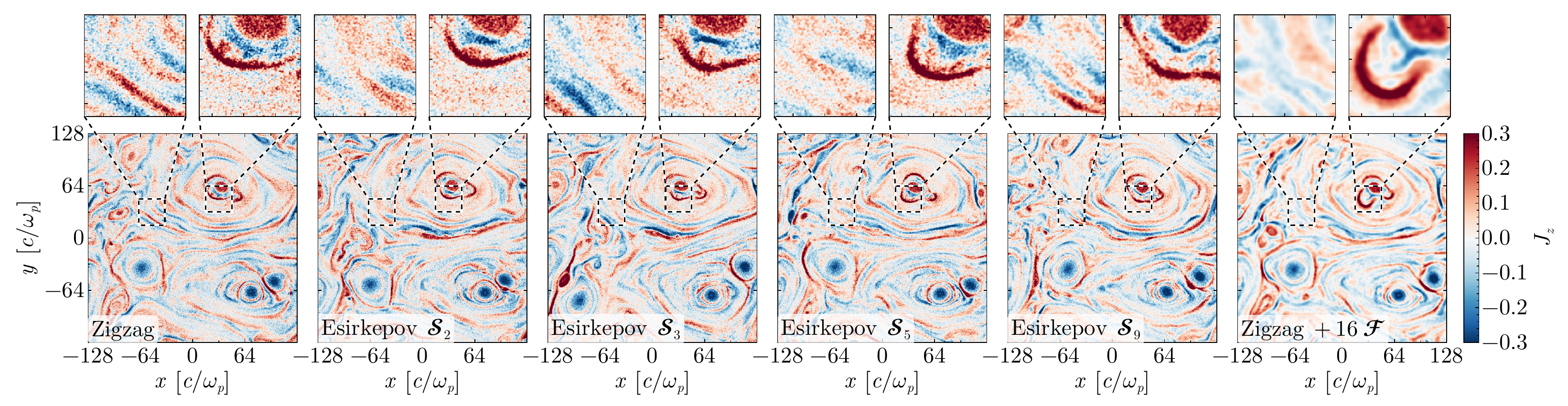}
  \caption{$J_z$ component in the final state of the decaying turbulence test after three Alfvén crossing times. From left to right, we show the runs with \zz deposit, \es deposit with 2\thu, 3\rd, 5\thu, and 9\thu~order shape functions. The right-most panel shows the test using the \zz deposit, using 16 current filter passes.}
  \label{fig:decaying_turb_images}
\end{figure*}

In this section, we investigate the performance of higher-order shape functions in a decaying turbulence setup, representing a standard study case for astrophysical plasmas \citep[e.g.][]{Zhdankin2017, Comisso2018}. We focus on a 2D case, for which we initialize an in-plane magnetic field perturbation $\delta \bm{B} =\{\delta{B}_x, \delta{B}_y\} $ as

\begin{equation}
  \delta \bm{B} = i\sum_k \delta \bm{B}_{\bm{k}} \frac{\bm{k}\times \bm{B}_0}{|\bm{k}\times{\bm{B}_0}|}\exp{\left[i(\bm {k}\cdot\bm{x}+ \phi_{\bm{k}})\right]},
\end{equation}

\noindent where $\delta \bm{B}_{\bm{k}}$ is an amplitude of a given Fourier mode with a wavevector $\bm{k}$, $\bm{B}_0$ is an out-of-plane guide field $B_z = |\bm{B}_0|$, and the random phases $\phi_{\bm{k}}$ satisfy the condition $\phi_{\bm{k}} = -\phi_{-\bm{k}}$ to ensure that $\delta \bm{B}$ is real. We excite wavevectors with components $k_{x,y} = 2\pi j_{x,y}/L$, where $j_{x,y}\in[1;4]$, and $L$ is the simulation box size, equal in both directions. The amplitudes $\delta\bm{B}_{\bm{k}}$ are identical for all wavenumbers, and we ensure that $|\delta \bm{B}|/|\bm{B}_0|=1$ at the start of the simulation. We start with initially hot Maxwellian pair plasma with $T_0= 0.49 \: m_e c^2$ and magnetization $\sigma_0 \equiv B_0^2/(4\pi n_0 m_e c^2) = 1$ (with $n_0$ being the total number density). In the base case, the simulation is resolved with $4$ cells per plasma skin depth, $d_e = 4\Delta x_0$, and the simulation box size is kept at $L= 256 \: d_e$. Below, we also present a series of runs with increased resolution: $\Delta x\in [1;1/2;1/4;1/8]\Delta x_0$. We evolve the system for a total of three Alfv\'en-crossing times $t_{A} = L/v_{A}= L\sqrt{(\sigma_0+1)/{\sigma_0}}/c$ to allow for the turbulent cascade to fully develop. The random phases $\phi_{\bm{k}}$ are initialized with the same seed in every run to ensure reproducibility.

We show a visualization of the final state of $J_z$, the out-of-plane current component, in Fig.~\ref{fig:decaying_turb_images}. From left to right, we show the simulation with the \zz deposit, and the \es deposit using 2\nd, 3\rd, 5\thu, and 9\thu-order shape functions, as well as the \zz deposit with the addition of 16 current filter passes. Generally, we see a gradual decrease in background noise, while maintaining the fine-scale structure of eddies and current layers in-between, when moving from \zz to higher order shape functions. Including current filtering further reduces the background noise.

\begin{figure}
  \includegraphics[width=0.95\linewidth]{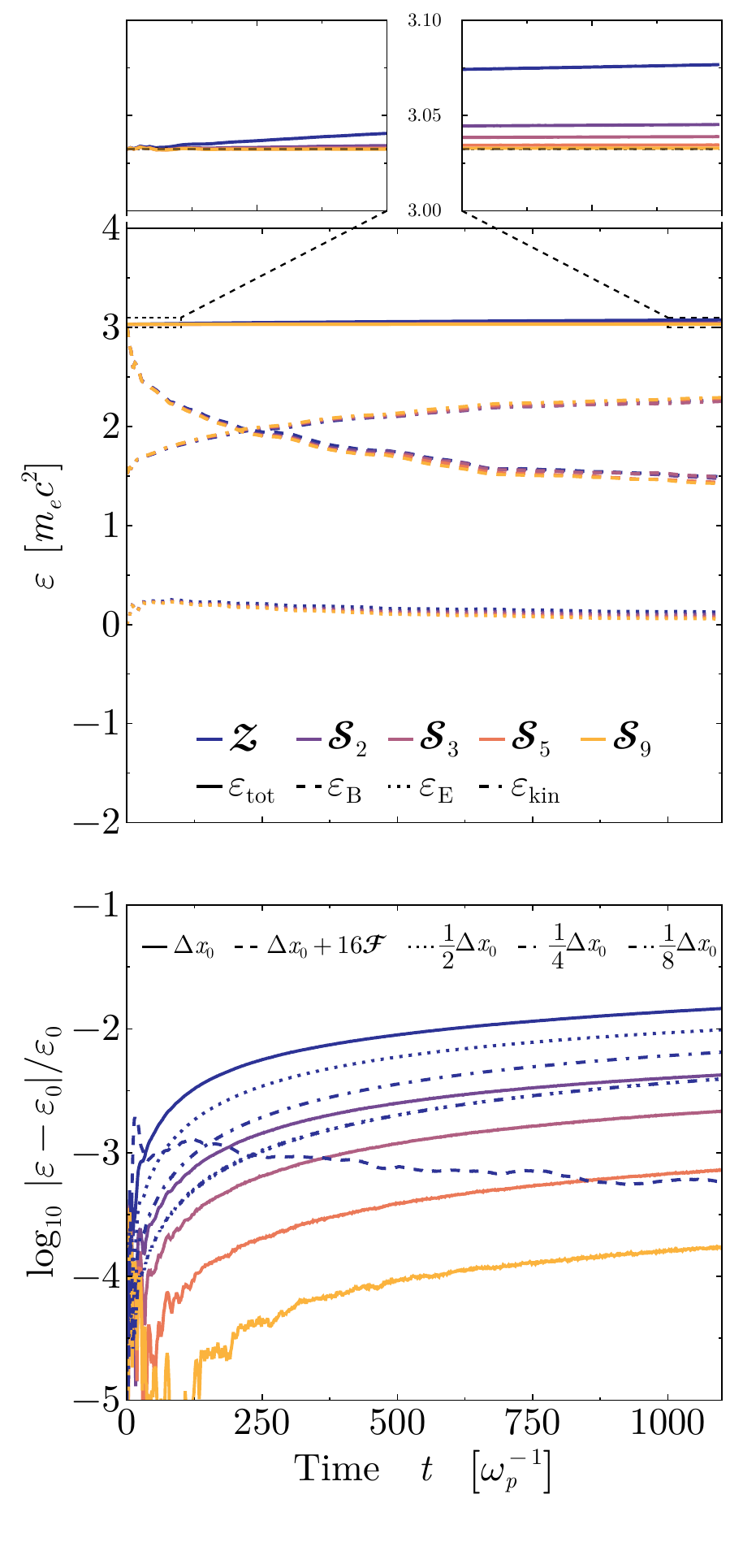}
  \caption{Energy components in the decaying turbulence test as a function of time. The colors refer to \zz deposit ($\mathcal{Z}$), \es deposit using 2\nd, 3\rd, 5\thu, and 9\thu~order shape functions ($\mathcal{S}2 -\mathcal{S}_9$). \textit{Top panel:} Individual energy components for total ($E_\mathrm{tot}$), magnetic ($E_B$), electric ($E_E$) and kinetic ($E_\mathrm{kin}$) energy. The zoom-ins magnify the total energy at the beginning and the end of the simulation. The dashed line indicates $E_\mathrm{tot}$ at the beginning of the simulation. \textit{Bottom panel:} Relative energy error of the total energy as a function of time. Colors indicate the same runs as in the top panel. For the \zz~runs we show the default resolution in solid lines, a run at the default resolution with 16 current filter passes in dashed lines, and runs at $2\times$ (dotted), $4\times$ (dash-dotted), and $8\times$ (dash-dot-dotted) the default resolution.}
  \label{fig:decaying_turb_energy}
\end{figure}

Fig.~\ref{fig:decaying_turb_energy} shows the total energy in the system per particle, $\varepsilon$, in the top panel, and the relative change in total energy over time in the bottom panel. We decompose the total energy in the top panel $\varepsilon_\mathrm{tot}$ (solid lines) into separate magnetic and electric field contributions, $\varepsilon_\mathrm{B}$ (dashed lines), and $\varepsilon_\mathrm{E}$ (dotted lines), and the kinetic energy of particles, $\varepsilon_\mathrm{kin}$ (dash-dotted lines). Small panels above show zoom-ins on the total energy at the beginning and end of the simulation. The bottom panel shows the logarithm of the relative total energy change over time. Colors indicate the different runs introduced above.

We find a clear convergence trend when moving from \zz to higher-order shape functions. From \zz to \es 2\nd-order, the relative error of the total energy at the end of the simulation decreases by almost half an order of magnitude. This convergence trend continues, with the 3\rd-order reducing the relative error by another quarter of magnitude, with the 5\thu-order reducing the error by another half order of magnitude. With the 9\thu-order, the error at the end of the simulation is reduced by two orders of magnitude compared to the standard \zz deposit.

To study the impact of the grid resolution (number of cells per skin depth) on these results, we performed three simulations using the \zz~deposit at $\times2$, $\times4$, and $\times8$ the base resolution. These tests are shown in the bottom panel of Fig.~\ref{fig:decaying_turb_energy} with dark blue dotted, dash-dotted, and dash-dot-dotted lines, respectively. As expected, we find improved energy conservation for higher resolution. However, it showcases the superior energy conservation of high-order shape functions, since the error in total energy at $\times8$ the base resolution for \zz~is of the same order as the 2\nd-order \es with base resolution.

\begin{figure}
  \includegraphics[width=\linewidth]{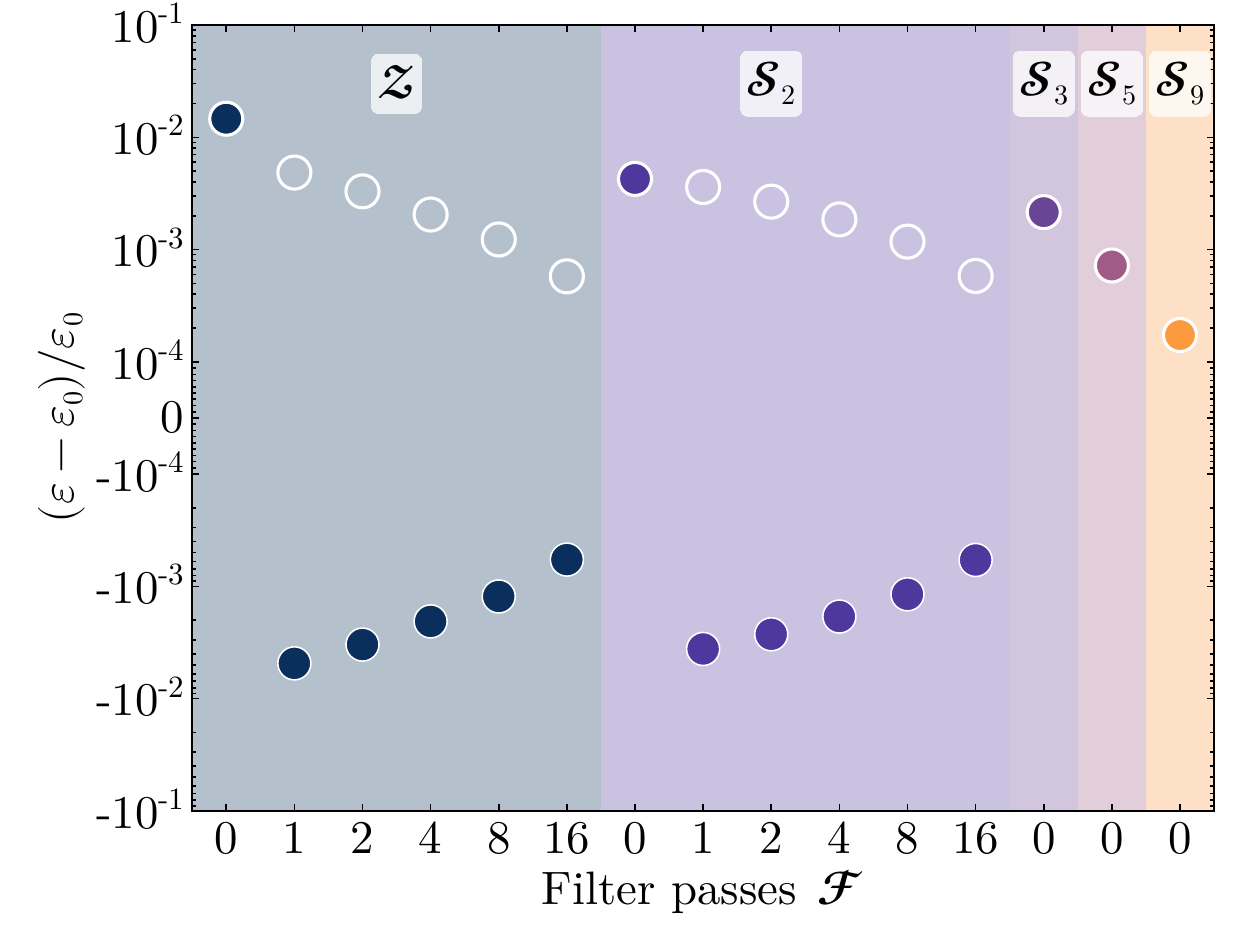}
  \caption{Convergence behavior of energy conservation in the decaying turbulence test. The color of the dots and the band corresponds to the indicated current deposition method. Each dot corresponds to a run with the number of current filter passes shown on the x-axis. Solid dots show the real value of the error, white circles the absolute value.}
  \label{fig:decaying_turb_energy_convergence}
\end{figure}

\begin{figure}
  \includegraphics[width=\linewidth]{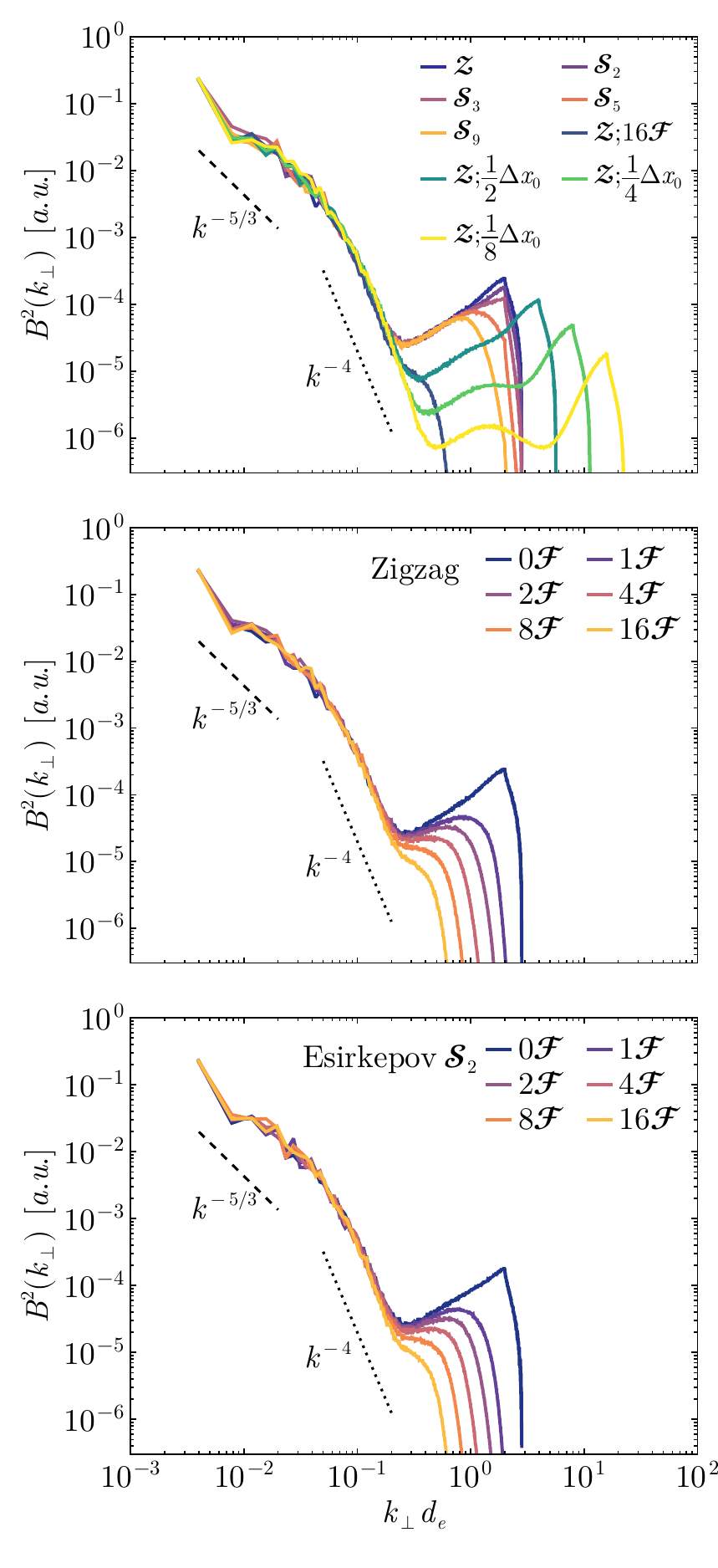}
  \caption{Magnetic power spectra for the decaying turbulence test. \textit{Top panel:} Spectra for the differen deposit schemes \zz, and \es 1\st-9\thu-order, compared with \zz using 16 current filter passes, and \zz at 2x-8x base resolution. \textit{Middle panel:} Spectra for \zz using 0-16 filter passes. \textit{Bottom panel:} 2\nd-order \es deposit using 0-8 filter passes.}
  \label{fig:decaying_turb_spectrum}
\end{figure}

Including 16 current filter passes when using the \zz~deposit also results in improved energy conservation, with the error level comparable to using the 5\thu-order shape function with no filters. However, this only relates to the absolute value of the error, as the current filtering changes the convergence behavior. While high-order shape functions stabilize against numerical heating, i.e., they suppress a growth in total energy, and the energy conservation converges from above, the inclusion of current filtering leads to a convergence from below, i.e., energy is lost over the course of the simulation. To illustrate this behavior, we show the energy error between the beginning and the end of the simulation in a pseudo-log scale in Fig.~\ref{fig:decaying_turb_energy_convergence}. The y-axis shows the energy error, while the x-axis indicates the number of current filter passes for each deposit method. Runs using different deposit methods are indicated with the same color of marker and background, and are labeled in the center of the background band. We performed the same runs with \zz and 2\nd-order \es using $0$, $1$, $2$, $4$, $8$, and $16$ filter passes, respectively. For comparison, we also show 3\rd, 5\thu, and 9\thu-order without filter passes. Solid dots indicate the real error, while white circles show the absolute error, to guide the eye on the convergence trend.

In all runs without filter passes, we observe the convergence trend outlined above. As soon as filtering is included, the convergence trend changes, and the final energy of the system is below the initial energy. This energy loss decreases in magnitude with more current filter passes. Ultimately, purely from the energy conservation standpoint, the \zz deposit with $2$ passes is comparable to the 2\nd~order \es with no filters, while the \zz deposit with $16$ passes is comparable to the 5\thu~order \es with no filters.

We find that this convergence trend is caused by the filtering of small-scale noise.
To illustrate this, we show the magnetic power spectra of the runs in Fig.~\ref{fig:decaying_turb_spectrum}.
The top panel shows the spectra for all simulations shown in Fig.~\ref{fig:decaying_turb_energy} from dark blue to orange, and simulations with 16 filter passes or increasing resolution from dark teal to yellow.
In the middle panel, we show the spectra for simulations using a \zz deposit with an increasing number of current filter passes, while the bottom panel shows the same for a 2\nd-order \es deposit.
All spectra are normalized to the same maximum value to facilitate direct comparison.

Generally, we find that the global power spectra are not affected by the choice of shape function or the number of current filter passes.
The difference arises in the noise-dominated part of the spectrum at large $k_\perp d_e \geq 0.3$, where higher-order shape functions or filters remove small-scale noise.
In regimes where extracting physical meaning from these scales is questionable, higher-order shape functions provide less benefit than current filtering, since filtering is computationally cheaper.
In simulations where information of high-$k$ modes could be relevant (e.g., electron-ion turbulence), the improved energy conservation and the intricate interaction of high-$k$ modes with electrons may lead to a benefit of using higher-order shape functions with fewer filter passes.
The same holds for 3D simulations, where a large number of filter passes can have a significant impact on performance. In this scenario, balancing the number of filter passes with the interpolation order may optimize code performance and improve results.

\subsection{Field solver\label{sec:convergence_fielstencil}}

In this section, we summarize the tests for the improved stencil in the Faraday solver, introduced in Sec.~\ref{sec:methods_fieldsolver}.

\subsubsection{Single particle}

\begin{figure*}[t]
  \centering
  \includegraphics[width=\linewidth]{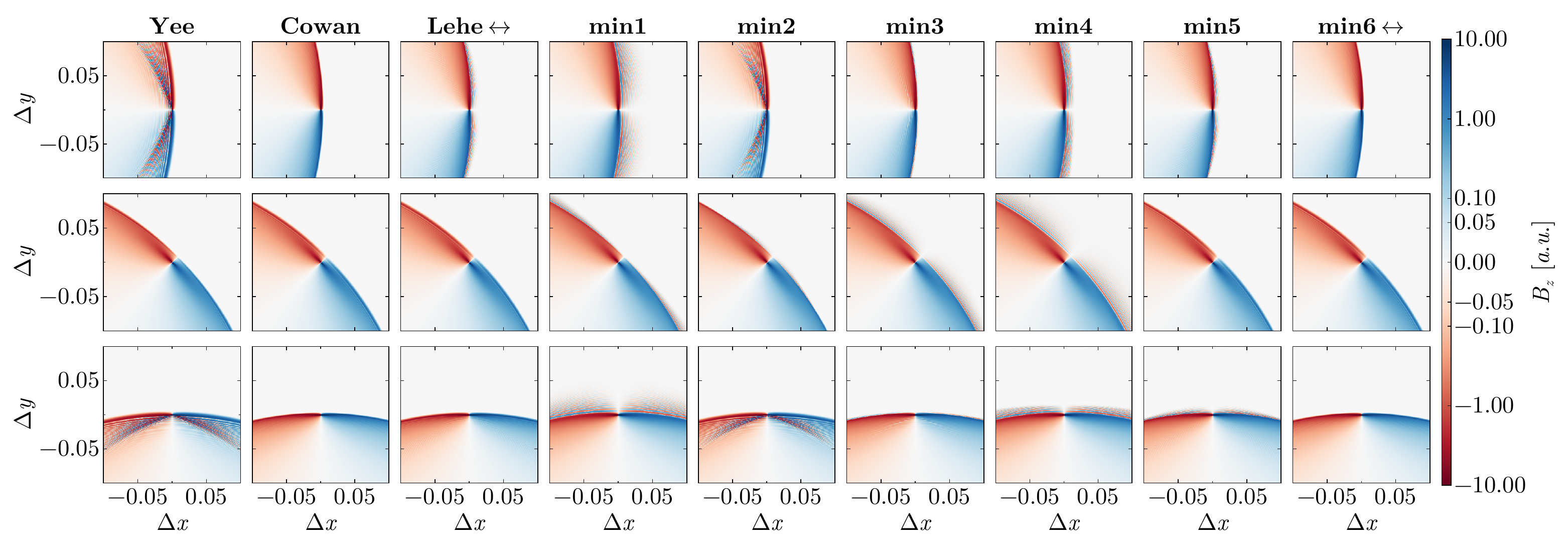}
  \caption{$B_z$ induced by the motion of a single electron moving in vacuum with $\gamma = 10$. From top to bottom, rows correspond to an inclination of the particle's propagation direction of $0^\circ$, $45^\circ$, and $90^\circ$. The columns are titled according to the stencil used in the Faraday solver. The numerical Cherenkov effect manifests as additional waves ahead or behind the particle.}
  \label{fig:blinne_fig04}
\end{figure*}

We first reproduce the results by \citet{Blinne2018}, specifically their fig.~4. We set up a single electron in vacuum with $\gamma = 10$ moving at angles of $0^\circ$, $45^\circ$, and $90^\circ$ with respect to the $x$-axis.\footnote{Note that because we only initialize a single negatively charged particle, with zero external fields, the requirement $\nabla\cdot\bm{E}=4\pi\rho$ enforces a phantom charge to exist indefinitely in the same position where our injected particle has been initialized.} All the simulations are performed with the standard \zz deposit, using 8 filter passes. The results can be seen in Fig.~\ref{fig:blinne_fig04}, where we show the $B_z$-component of the particle's magnetic field. Rows show the different inclinations, while columns show the runs with the different field stencils as indicated in the titles. We plot $B_z$ in pseudo-log scale to over-emphasize small fluctuations, especially ahead of the wave front.

\begin{itemize}
  \item The standard \texttt{Yee} stencil shows strong numerical artifacts in the wake of the wave front for the inclinations of $0^\circ$ and $90^\circ$. This is a key indicator for numerical Cherenkov radiation, introduced by an artificially slower speed of the electromagnetic wave.
  \item The \texttt{Cowan} solver remains remarkably clean in all inclinations. This is especially advantageous, since the CFL condition is chosen to be very large, reducing computational cost.
  \item In contrast to that, the \texttt{Lehe} solver introduces unphysical fluctuations in front of the wave for the $0^\circ$ inclination. These fluctuations are not present in the other inclinations, however.
  \item \texttt{min1} similarly shows these fluctuations, but in all inclinations.
  \item \texttt{min2} behaves similarly to the Yee solver, but also shows upstream fluctuations in the $45^\circ$ case.
  \item \texttt{min3} is cleaner in $0^\circ$ and $90^\circ$, but shows fluctuations in $45^\circ$.
  \item \texttt{min4} shows upstream fluctuations ahead of the wave front only, while the fields behind the wave front are clean. Given the considerably lower CFL condition, we find this solver to be the least convenient for most test cases.
  \item \texttt{min5} shows some fluctuations both ahead and behind the wave front; however, the fluctuations are considerably less pronounced than in the previous min stencils.
  \item Finally, \texttt{min6} remains the most stable under all inclinations. In combination with the large CFL this stencil is optimized for, it provides an excellent optimization for problems that allow for large timesteps.
\end{itemize}

\subsubsection{Drifting plasma}

\begin{figure*}
  \centering
  \includegraphics[width=\linewidth]{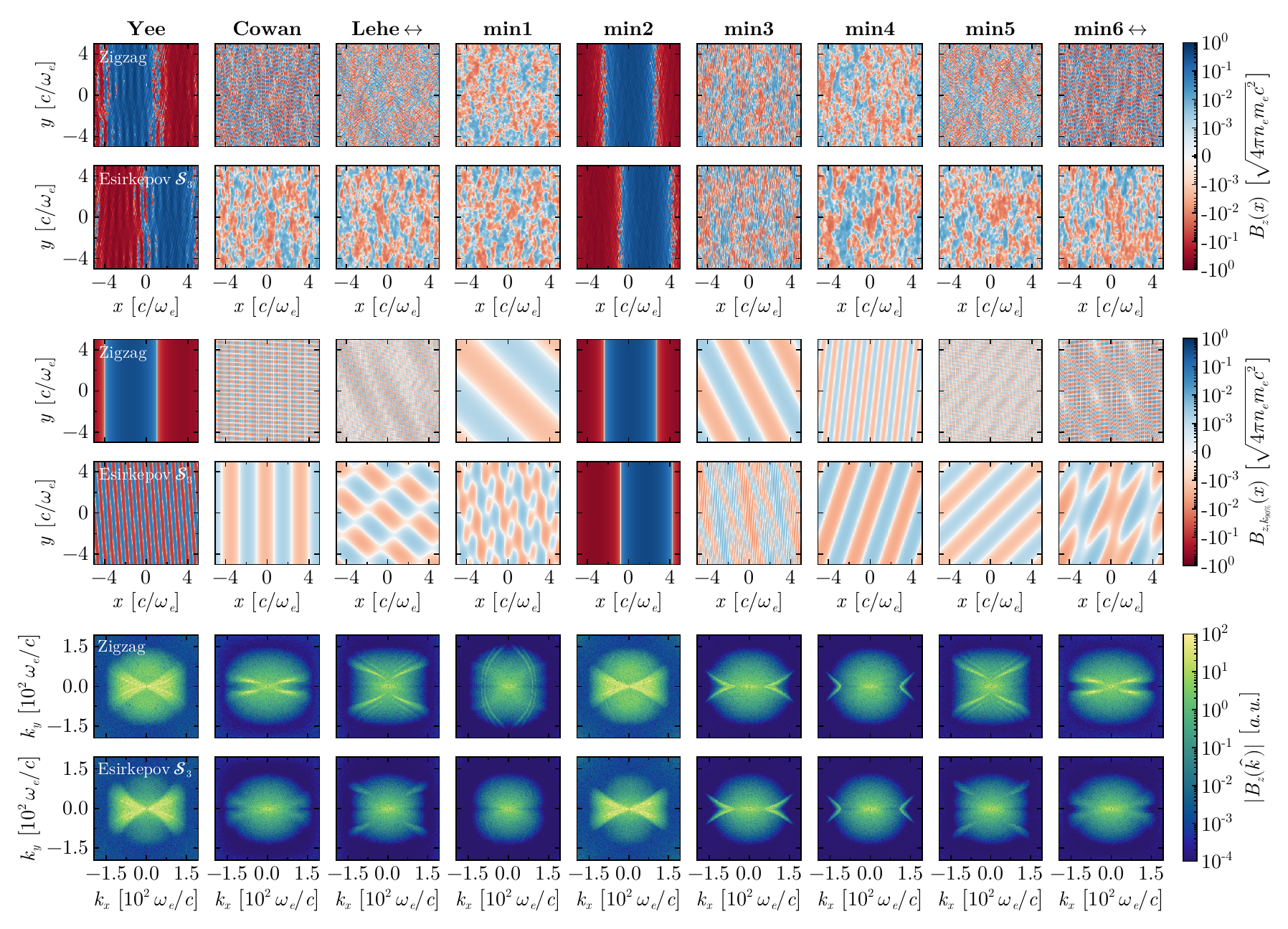}
  \caption{Final state at $t = 10^4 \: \omega_p t$ of a warm plasma with $E = 0.08 m_e c^2$ drifting with $\gamma = 10$ in a periodic setup. From left to right, we show the results for all field stencils tested in \citet{Blinne2018}. \textit{Top panels:} Absolute value of the out-of-plane magnetic field component ($B_z$). \textit{Middle panels:} $B_z$-component, after applying a high-pass filter in Fourier space to filter out the modes containing less than 90\% of the energy. \textit{Bottom panels:} Fourier transform of the top panels $B_z(\hat{k})$. We show the results with the standard \zz deposit in the top rows, while we use a \trd-order \es deposit in the bottom rows.}
  \label{fig:fieldstencil_drifting_plasma}
\end{figure*}

\begin{figure}
  \includegraphics[width=\linewidth]{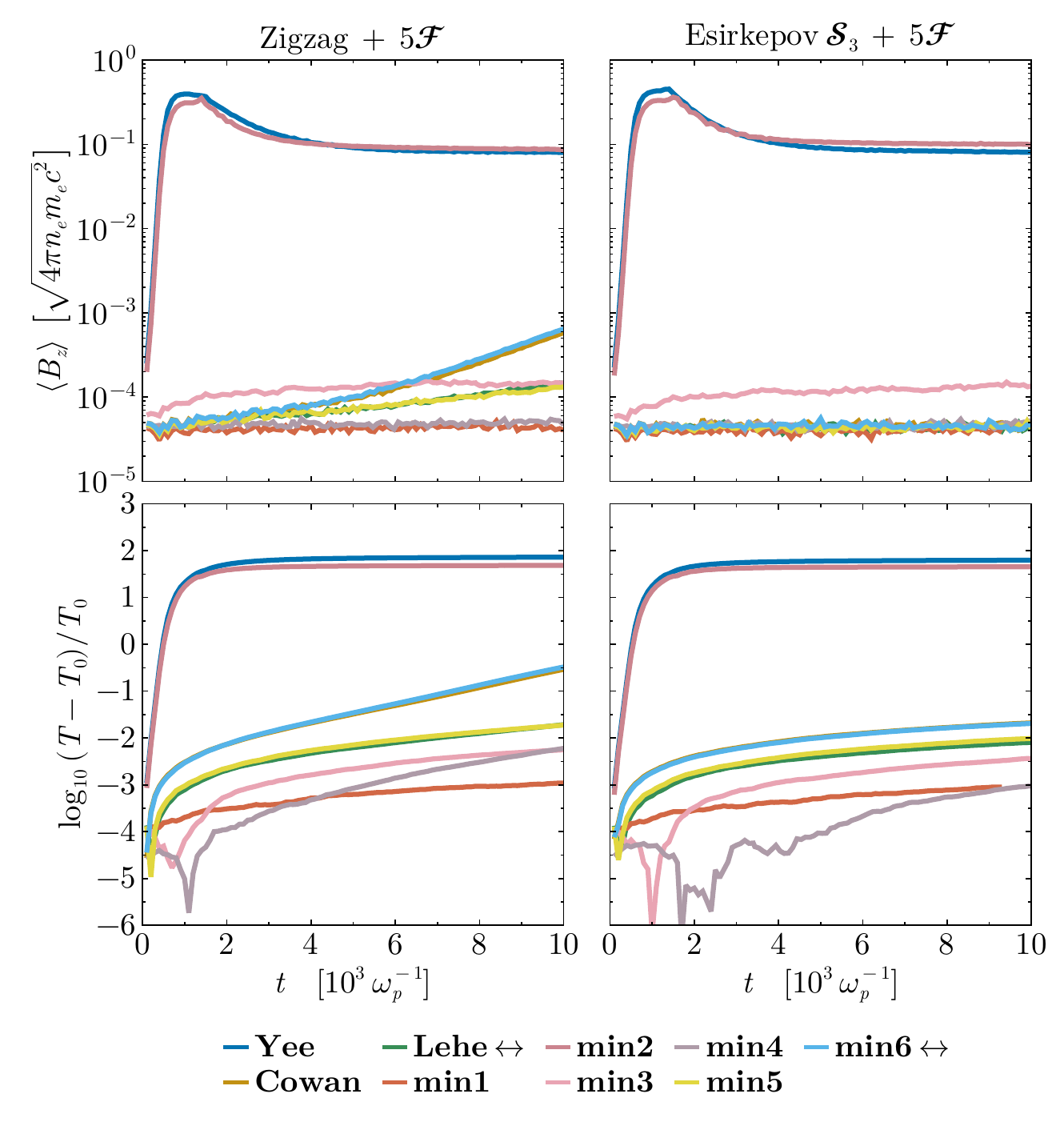}
  \caption{Time evolution of the drifting plasma test shown in Fig.~\ref{fig:fieldstencil_drifting_plasma}. Colors indicate the different field stencils. \textit{Top panels:} Mean of the absolute value of the $B_z$ component. \textit{Bottom panels:} Relative temperature increase of the plasma. \textit{Left panels:} Using the standard \zz deposit. \textit{Right panels:} Using the \es deposit with a 3\rd~order shape function.}
  \label{fig:stencil_heating}
\end{figure}

\begin{figure}
  \centering
  \includegraphics[width=0.95\linewidth]{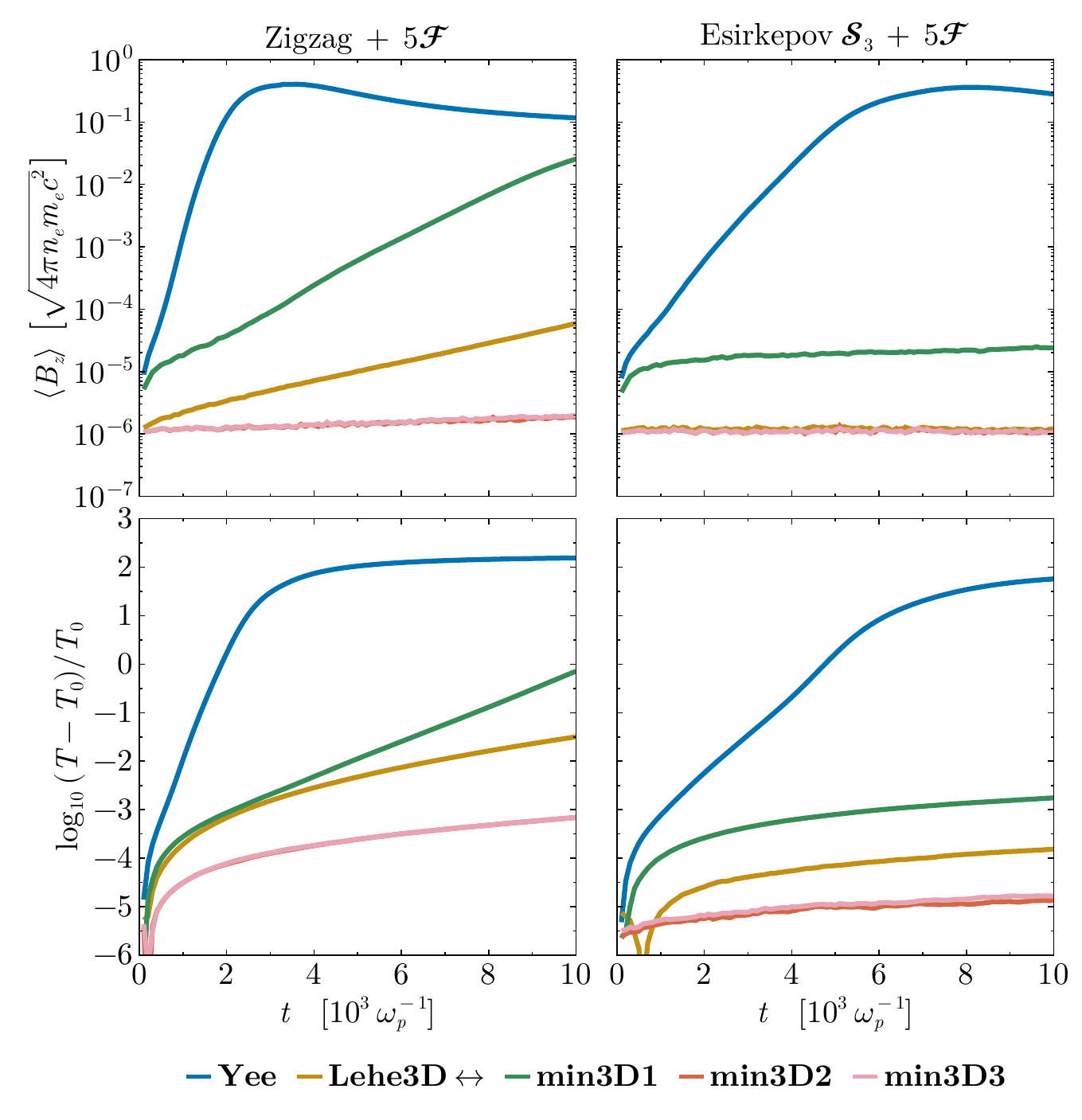}
  \caption{Similar to Fig.~\ref{fig:stencil_heating} we show selected field stencils optimized for 3D simulations.}
  \label{fig:stencil_heating_3D}
\end{figure}

For a more realistic and quantitative test case, we set up pair-plasma with an initial temperature $T = 0.08 \: m_e c^2$ and a drift along the $+\hat{\bm{x}}$ direction with a bulk $\gamma = 10$, on a 2D plane with periodic boundary conditions. We resolve the skin-depth, $d_e\equiv c/\omega_e$, with $25.6$ cells to avoid any numerical heating due to an under-resolved Debye length, as in the test setup of Sec.~\ref{sec:heating}. For this test, we use $5$ current filter passes, a relatively conservative number \citep[compared to the same tests performed with $20$ filter passes in][]{Groselj2022}, to emphasize the impact of the field stencils. We expect the fast bulk motion of plasma to develop a strong Cherenkov instability, which will manifest itself as a growth of $B_z$ and result in considerable heating of the plasma.

The results of this test are shown in Fig.~\ref{fig:fieldstencil_drifting_plasma}.
The two top rows show the 2D plot of $B_z$, while the lower two panels show the absolute value of the Fourier transform of $B_z$. The middle two panels show the same data as the top two panels, namely, the $B_z$-component, but normalized to its maximum value and after applying a high-pass filter in Fourier space that captures the modes containing more than 90\% of the total energy. Each column is labelled according to the field stencil used in the test. The respective top rows in each group use the standard \zz deposit, while the bottom rows use a 3\rd-order \es deposit, as labeled. We also show the time evolution of the magnitude of $B_z$ and the relative temperature increase in the top and bottom rows of Fig.~\ref{fig:stencil_heating}, respectively.

We want to first focus on the \zz deposit. The baseline for this test is the performance of the standard \texttt{Yee} stencil. At the end of the test, we find a significant increase in $B_z$, dominated by a long $\bm{k}\parallel \hat{\bm{x}}$ mode, and higher frequency diagonal modes. The increase in $B_z$ by $1.5$ orders of magnitude (dark blue line in the top-left panel of Fig.~\ref{fig:stencil_heating}) leads to a numerical increase of the plasma temperature by almost two orders of magnitude (dark blue line in the bottom-left panel of Fig.~\ref{fig:stencil_heating}). Within one fourth of the total runtime, the growth of the instability saturates, leading to a stable, but noisy, new configuration by the end of the run.

In direct comparison, the \texttt{Cowan} stencil performs better, leading to less numerical heating, weaker $B_z$, but strongly pronounced diagonal modes in $B_z$, visible in the second panel of the top-most row in Fig.~\ref{fig:fieldstencil_drifting_plasma}.

The \texttt{Lehe} stencil does not fulfill Eq.~\eqref{eq:stencil_divB} and is explicitly optimized to suppress the Cherenkov instability along the x-direction. This strongly suppresses the growth of $B_z$, and numerical heating (Fig.~\ref{fig:stencil_heating}), and the final state is dominated by much weaker $B_z$ fluctuations at high $k$ values diagonal to the propagation direction.

\texttt{min1} performs the best out of the stencils not fulfilling Eq.~\eqref{eq:stencil_divB} and not being optimized for a preferential direction. Numerical heating and $B_z$ increase remain minimal throughout the test run, and the fluctuations have smaller modes (higher $k$ values).

The \texttt{min2} stencil has been constructed to fulfill~\eqref{eq:stencil_divB}, and is optimized for a very large CFL. This leads to a result very similar to the \texttt{Yee} stencil. Both the increase in $B_z$ and the numerical heating are almost identical (purple line in Fig.~\ref{fig:stencil_heating}), and the final state is dominated by a strong low-frequency mode along $\hat{\bm{x}}$.

The stencils \texttt{min3} and \texttt{min4} both fulfill~\eqref{eq:stencil_divB}, are not optimized for a preferential direction, but are optimized for different values of CFL. They perform very similarly, with very little numerical heating and $B_z$ growth. Due to the moderate CFL of \texttt{min3}, it is a viable option for numerous setups that suffer from Cherenkov instability; however, \texttt{min4}, while performing marginally better, due to its considerably smaller CFL, can be much more computationally expensive.

The stencils \texttt{min5} and \texttt{min6} do not fulfill~\eqref{eq:stencil_divB}. As such, they behave very similarly to the \texttt{Lehe} and \texttt{Cowan} stencils, albeit with different directional behaviour. In this context, they provide an alternative with comparable CFL values.

The runs using a 3\rd-order \es deposit generally follow the trend of the runs using the \zz deposit. They are shown in the bottom rows of Fig.~\ref{fig:fieldstencil_drifting_plasma}, and the right column of Fig.~\ref{fig:stencil_heating}. A notable difference is that the 3\rd-order runs filter out more of the smaller wavenumber modes, leading to a general decrease in $B_z$ growth and numerical heating.
In all cases, the 3\rd-order tests produce similar or improved results for the numerical stability of the drifting plasma.

In Fig.~\ref{fig:stencil_heating_3D}, we show tests for the optimized stencils in 3D. To save computational time, we performed these tests in a smaller volume, with $128^3$ grid cells and half the transverse size, while keeping the same number of ppc and filter passes as in the 2D case.

Generally, we find that the extension to 3D retains similar stability improvements as we saw for 2D; however, the addition of higher-order shape functions provides stronger stability improvements in 3D than in 2D. This can be seen in the standard \texttt{Yee} stencil, where the simulation using a 3\rd-order \es deposit scheme shows a reduced growth rate of the numerical Cherenkov instability. While the \texttt{Lehe3D} stencil significantly reduces the numerical heating already for the \zz deposit, it is further reduced in the 3\rd-order simulation.
Our optimized stencils \texttt{min3D1}--\texttt{min3D3} perform better than the \texttt{Yee} stencil, however \texttt{min3D1} only retains considerable stability when paired with a 3\rd-order shape function.
\texttt{min3D2} provides significantly more stability, especially with the 3\rd-order deposit, where heating and and $B_z$-growth are almost completely suppressed.
\texttt{min3D3} performs identically to \texttt{min3D2}, and with its larger target CFL, may be a good option to mitigate the Cherenkov instability in setups that allow such a large CFL number, although we encourage potential users to perform tests in their specific setup.

We have tested further stencils with larger target values for \texttt{CFL} with up to CFL=0.85, but find no further stable configurations for the stencil symmetry properties imposed.

\subsubsection{Relativistic shock}
\begin{figure*}[t]
  \centering
  \includegraphics[width=\linewidth]{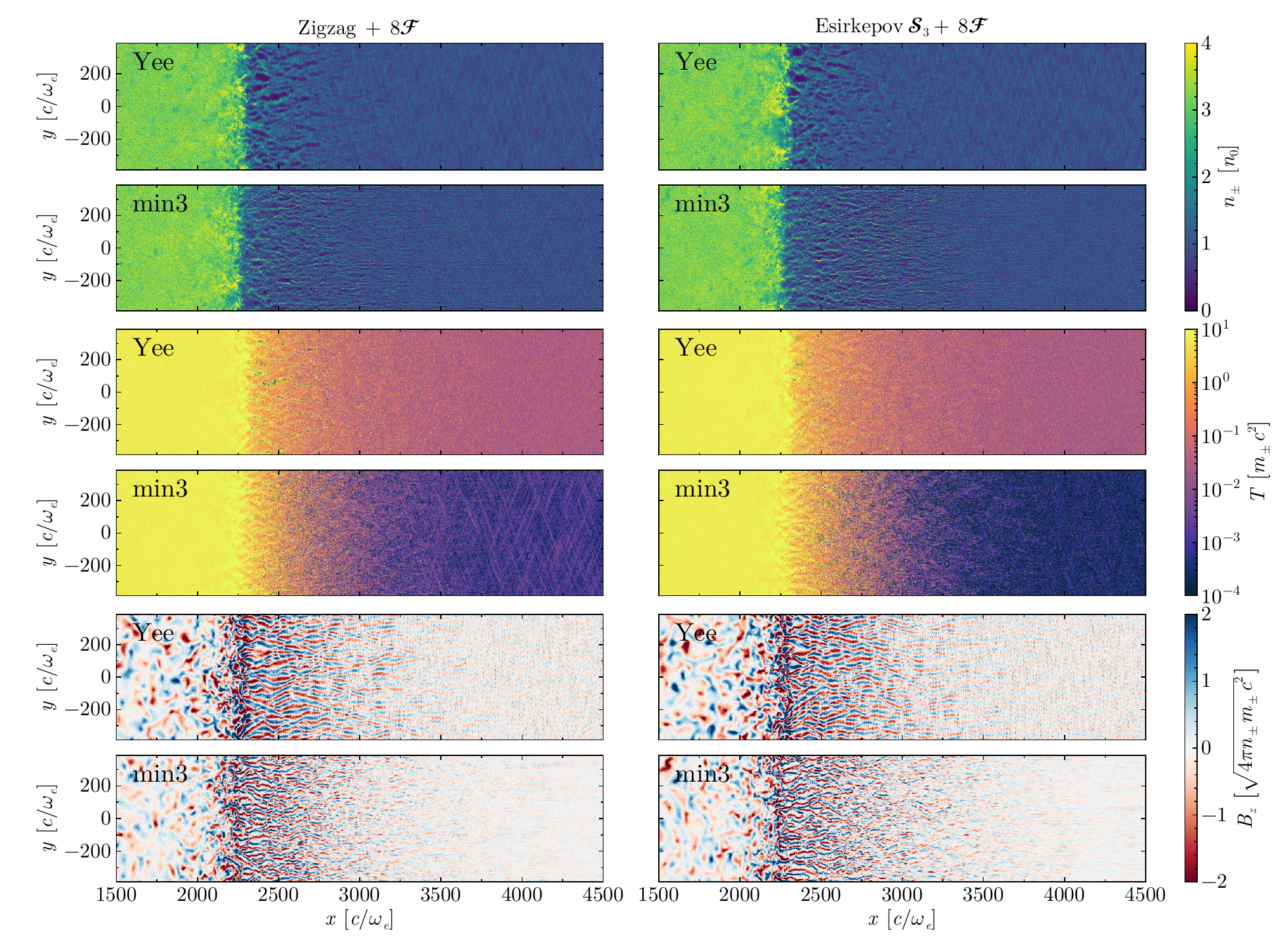}
  \caption{Result of the relativistic pair-plasma shock setup. \textit{Top panels}: Mass density. \textit{Middle panels:} Temperature of the plasma. \textit{Bottom panels:} Out-of-plane magnetic field component. For all grouped plots, the respective top panel shows the simulation employing the standard \texttt{Yee} stencil, while the bottom panel shows the \texttt{min3} stencil. The simulations in the left column were performed with the standard \zz deposit, while simulations in the right column were performed with the \es deposit using a 3\rd~order shape function.}
  \label{fig:rel_shock}
\end{figure*}
\begin{figure*}
  \centering
  \includegraphics[width=\linewidth]{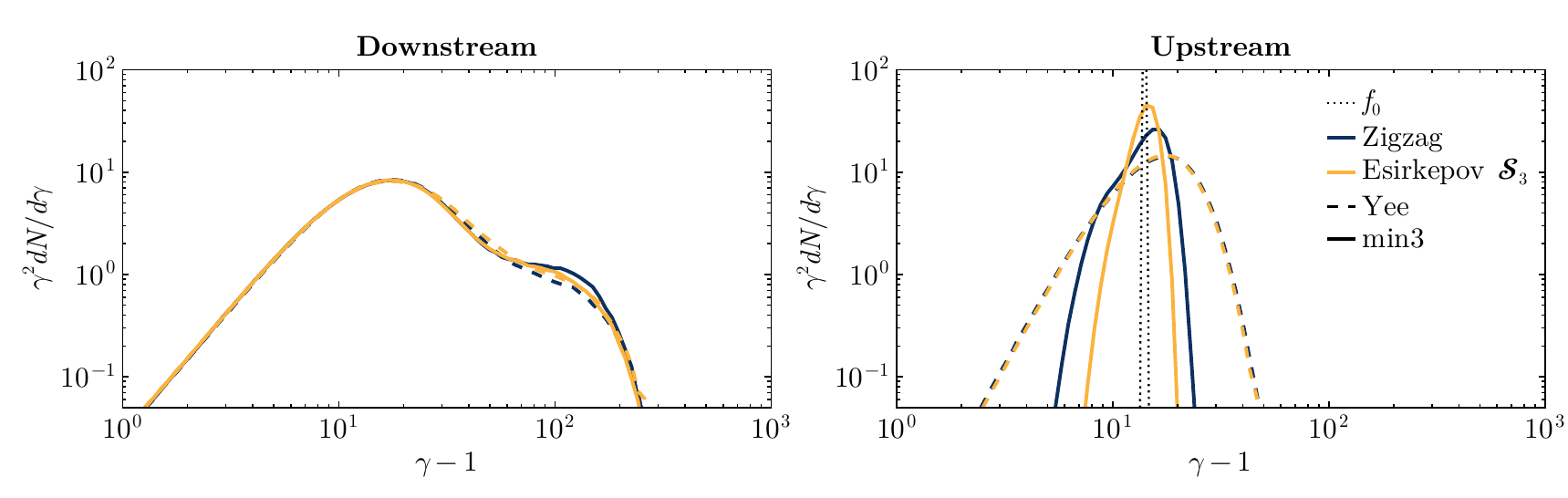}
  \caption{Spectra of electrons in the shock shown in Fig.~\ref{fig:rel_shock}. Blue lines correspond to the runs with \zz deposit, yellow lines to the runs with the \es deposit using a 3\rd~order shape function. Dashed lines show the spectra of the simulations with the \texttt{Yee} stencil, while the solid lines show simulations with the \texttt{min3} stencil. \textit{Left panel:} Downstream spectra selected in the range $x \in [1000, 2000] \: c/\omega_p$. \textit{Right panel:} Upstream spectra selected in the range $x \in [4000, 5000] \: c/\omega_p$.}
  \label{fig:rel_shock_spectra}
\end{figure*}

Finally, to study the impact of the choice of the Faraday stencil in a realistic problem, we repeat the example of an unmagnetized relativistic pair-shock in 2D presented in \citet{Hakobyan2025}. This shock setup consists of pair-plasma with a temperature $T = 10^{-4} \: m_e c^2$, drifting with bulk $\gamma = 10$ against a reflecting wall in the $-\hat{\bm{x}}$ direction. The wall at $x=0$ is a perfect conductor for the fields and reflective for particles. With this setup, the simulation is performed in the downstream frame of the shock. The skin-depth is resolved with $2.5$ cells, and the upstream density is sampled with $8$ ppc: effectively 40\% of the resolution employed in \citet{Spitkovsky2008}. Each simulation is performed with $8$ current-filter passes and the same $c\Delta t=0.5 \Delta x$.

We show the results of the simulations in Fig.~\ref{fig:rel_shock}.
Top-row panels show the total pair-density, the middle panels show the temperature of the plasma, computed from $T = \mathcal{T}_{22}/n_0$, where $\mathcal{T}_{22}$ is the $22$-component of the energy-stress tensor, and the bottom-row panels show $B_z$. The simulations in the left column were performed with the standard \zz deposit, while the simulations in the right column were performed with the 3\rd-order \es~deposit. Within each block, the top panel shows the simulation using the \texttt{Yee} stencil, while the bottom simulation uses the optimized \texttt{min3} stencil.

The numerical Cherenkov instability is very pronounced in both simulations using the \texttt{Yee} stencil, with barely discernible difference between \zz and 3\rd-order \es deposit. This can be seen in the far-upstream of the shock $x > 3000 \: c/\omega_e$, where diagonal modes in the density develop. It further manifests in $B_z$, where strong fluctuations form far upstream at the same distance from the shock.
The numerical heating caused by this instability is clearly visible in the temperature, which increases by almost three orders of magnitude in the far upstream, compared to the initialized value.

In contrast, when employing the \texttt{min3} stencil, this instability upstream is considerably suppressed. While the \zz~deposit still shows some Cherenkov instability further upstream, $x > 4000\: c/\omega_e$, it is almost entirely absent in the run using the 3\rd-order \es~deposit.

The difference between these runs mainly manifests in the region just ahead of the shock ($x \in [2250; 2750] \: c/\omega_p$), commonly referred to as the shock ramp. For the Yee stencil, numerical heating imparted by the Cherenkov instability leads to heating of the background, relativistically drifting plasma up to mildly relativistic temperatures ($k_{\rm B} T = 10^{-1} m_{\rm e} c^2$). Consequently, current filaments grow on the scale of the background plasma skin depth, increased by its specific enthalpy $h_{\rm p}$, $c/\omega_{\rm p} \propto h_{\rm p}^{1/2}$  \citep{Ohira2025}. As a result, artificially larger current filamentary structures develop within the shock precursor.

This numerical heating and its impact can also be seen in Fig.~\ref{fig:rel_shock_spectra}, where we show the energy distribution of particles both downstream ($x \in [1000; 2000] \: c/\omega_p$) -- left panel, and upstream ($x \in [4000; 5000] \: c/\omega_p$) -- right panel. In blue we show the runs using the \zz deposit, in yellow we show the runs using the 3\rd-order \es deposit. Dashed lines correspond to the runs with the standard \texttt{Yee} stencil, solid lines the \texttt{min3} stencil. For reference, we plot the spectrum of the initial particle distribution upstream with a black dotted line.

We clearly see a strong suppression of numerical heating upstream when using the \texttt{min3} stencil as compared to the \texttt{Yee} stencil. This is further suppressed when using the 3\rd order \es~deposit, instead of the \texttt{Zigzag}. Numerical heating in the \texttt{Yee} stencil runs results in the aforementioned over-heating of the shock ramp and changes the acceleration behavior, which manifests in a steeper power law of the spectrum downstream.

We note that in order to save on computational cost, we ran the simulation for a shorter time, and thus the power-law is not as well developed as in \citet{Hakobyan2025}. We therefore omit discussions of the long-term evolution of the shock, as this example should only provide a demonstration for a real-life application of the improved field stencil.

\section{Performance}
\label{sec:performance}

\subsection{Shape functions}
\begin{figure*}[t]
  \centering
  \includegraphics[width=\linewidth]{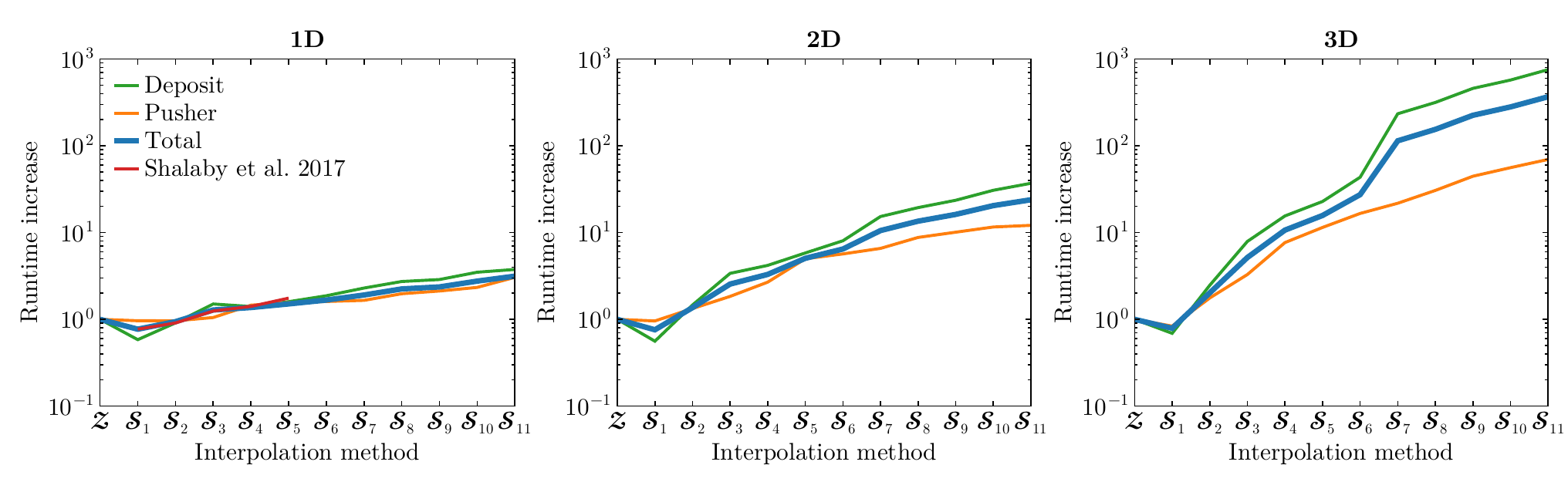}
  \caption{Runtime increase as a function of interpolation method. Here $\mathcal{Z}$ refers to the standard \zz deposit, while $\mathcal{S}_1 - \mathcal{S}_{11}$ refers to \es deposit with 1\st-11\thu~order shape functions. For reference, we plot the performance impact on the deposit and pusher reported by \citet{Shalaby2017} in their 1D simulations.}
  \label{fig:scaling}
\end{figure*}
We show the impact on runtime of using higher-order shape functions in Fig.~\ref{fig:scaling}.
The cost increase is split into current deposit (green), particle pusher (orange), and total cost increase (blue).
The cost increase was calculated using the Weibel test setup, as it is numerically simple, uses a constant number of particles, and is not strongly affected by load imbalance.
All tests were performed on a single tile of an Intel 1550 Max GPU and use $5.4\times 10^8$ particles per tile, constituting to 75\% of the available memory.
This assures that our tests are compute/memory limited by deposit and pusher.

In 1D, the general performance impact is insignificant.
We find that the deposit increases at maximum by a factor of 3.7 when employing an 11\thu-order shape function and the total runtime increases by a factor of 3.1.
As a reference, we show the scaling reported by \citet{Shalaby2017} for their implementation of higher order shape functions up-to 5\thu-order.
They find a cost increase for deposit and pusher of a factor of three between 1\st~and 5\thu-order, in line with our findings.

In 2D, the performance impact of using the \es deposit, compared to \zz, is more considerable.
First, we see a cost decrease when using the 1\st~order shape function with the \es deposit.
We attribute this to the smaller number of operations required in the deposit, since no mid-point construction in the particle trajectory is performed.
2\nd-order increases the total cost by a factor $\sim1.36$ and 3\rd-order by a factor $\sim2.5$ with respect to \zz.
When using 11\thu-order, we find a total runtime increase of a factor $\sim24$.

For 3D, the cost increase quickly becomes untenable.
While we again see a slight cost decrease for the 1\st-order \es deposit, 2\nd-order already increases the total computational cost by a factor of $\sim 2$, and 3\rd-order by almost a factor of $\sim 5$.
The considerable jump in cost-increase between 6\thu, and 7\thu-order is caused by the size of the cache, and therefore is architecture-dependent.
For GPUs with smaller caches, this jump happens at a lower order.
This is, however, of little relevance in production applications for current architectures, since the cost increase already surpasses one order of magnitude at this point and reaches two orders of magnitude for 11\thu-order.

Weak scaling performance is generally not affected by the choice of shape function order. However, due to the increased stencil size, a larger number of ghost cells is required when increasing the shape function order.
This impacts the communication time and, with that, the overall runtime.

We note that in 3D, depending on the deposition of the domain, current filtering can carry a significant fraction of the cost per timestep. It can therefore be beneficial to reduce the number of current filter passes in favor of a higher-order shape function. In our 3D testing, moving from 1\st~to 2\nd-order shape functions, but reducing the number of current filter passes from eight to four, retained the same computational cost, but proved to be significantly more stable in cases with an under-resolved Debye length.

In general, the performance of the \es deposit scheme can be improved by reducing the number of grid cells the current is deposited into, by combining the \es and \zz schemes, as introduced by \citet{Steiniger2023}.
They report an improved performance by $\sim 10\%$ when using this method.
This additional performance improvement can be considered for future work.

\subsection{Field stencil}

The computational cost of the fieldsolver is of the order $\sim 1 \%$ of the timestep and therefore negligible.
Hence, we omit an explicit scaling test.
The main driver of the cost increase for the total simulation is the value of \texttt{CFL}, for which the field stencil is optimized.

Under the conservative requirement that equation~\eqref{eq:stencil_divB} is fulfilled, this constraint can be significant when choosing \texttt{min4}.
On the other hand, \texttt{min2} is optimized for a large \texttt{CFL}, but shows poor convergence behavior.
In production runs, we find a good compromise in \texttt{min3}.
In test cases where a larger \texttt{CFL} produces reasonable results, and Eq.~\ref{eq:stencil_divB} is not required, the \texttt{Cowan} and \texttt{min6} solvers will be the most performant.
In our testing we find \texttt{min3} to be a good compromise between stability and \texttt{CFL} constraint.

Regardless, we encourage the user to perform tests for their specific setup using different field stencils.

\section{Conclusion}
\label{sec:conclusion}

In this work, we summarized our efforts to implement higher-order methods for the current deposit, field solver, and field interpolation into the PIC code \entity.
We performed extensive tests to show the accuracy and stability of the new methods and evaluated the performance impact of choosing higher-order approaches in standard PIC problems.
While none of these methods provides a silver bullet to tackle all numerical challenges in PIC simulations, they each have their advantages in specific cases.
Our findings can be summarized as follows:
\begin{itemize}
  \item The current deposit scheme by \citet{Esirkepov2001} as implemented in this work shows excellent charge conservation, with a weak dependence of the accuracy of the charge conservation on the order of the particle shape function.
  \item Higher order particle shape functions for current deposit and field interpolation considerably decrease the numerical stopping power of a plasma. Moving from 1\st~to 3\rd-order doubles the effective resolution, potentially allowing to offset the increased computational cost by employing fewer particles per cell.
  \item They further significantly reduce numerical heating and, with that, increase numerical stability when simulating cold plasmas. Moving from 1\st~to 2\nd-order decreases the required grid resolution to maintain a stable cold plasma by a factor $N\sim 4$, moving to 3\rd-order even by a factor of $N\sim 8$ per dimension. This can potentially reduce the total computational cost of a simulation by a factor of $N^{D+1}$, where $D$ is the dimensionality of the simulation.
  \item Including additional current filtering can further decrease numerical heating, equivalent to an additional factor of four in resolution increase.
  \item In realistic setups like CR streaming-induced instabilities, or decaying turbulence, higher-order shape functions reduce numerical noise. The global evolution and large-scale properties are, however, not significantly affected. Strongly subrelativistic regimes may nevertheless be more significantly affected.
  \item Higher-order shape functions considerably improve the energy conservation. Nonetheless, the same can be achieved with current filtering at a lower interpolation order.
  \item Specialized field stencils can considerably reduce the numerical Cherenkov instability, leading to more consistent single-particle fields, and reducing numerical heating in relativistically drifting plasmas.
  \item The choice of the field-stencil is very problem-dependent, and can be further improved by optimizing a stencil for a preferential spatial direction, if the problem has an appropriate symmetry.
  \item Combining higher-order field stencils with higher-order particle shape functions further reduces numerical heating, by stabilizing against small-scale noise.
  \item In simulations of relativistic shocks, this improves results by stabilizing the upstream plasma against the numerical Cherenkov instability.
  \item The performance impact of using higher-order shape functions is negligible in 1D, significant for $\mathcal{O}>5$ in 2D, and becomes prohibitive for $\mathcal{O}>3$ in 3D. This can be offset by the reduction in required numerical resolution for the same level of stability, but these offsets should be considered carefully to not reduce the physical validity of the simulation by under-resolving relevant processes.
\end{itemize}
These results lead us to make the following recommendations for the users:
\begin{itemize}
    \item Use 2\nd, or 3\rd-order shape functions whenever feasible.
    \item Offset the computational cost increase from moving to higher-order shape functions by reducing the number of current filter passes, or the number of particles per cell.
    \item For the best generality, use the \texttt{min3} optimized stencil in 2D simulations, and the \texttt{min3D2} stencil in 3D simulations.
    \item If a problem allows for large values of \texttt{CFL}, use the \texttt{Cowan} stencil in 2D, and the \texttt{min3D3} stencil in 3D. 
\end{itemize}
All of the presented changes to \entity~have been released in \texttt{v1.3.0}, and are publicly available. Timing and simulation results were obtained with \texttt{v1.4.0rc}.
Usage information can be found in the official documentation under \url{https://entity-toolkit.github.io/wiki/content/3-code/6-higherorder/}.
Parameter files, problem generators, and scripts to reproduce the results in this paper will be made available under \url{https://github.com/LudwigBoess/<ArXiV_ID>}.

\begin{acknowledgements}
  We thank Anatoly Spitkovsky, Sasha Philippov, Laurent Gremillet, Camille Granier, and Lorenzo Sironi for helpful discussions, and Nigel Tufnel for inspiring us to implement an $11^\mathrm{th}$ order shape function.
  This research was supported in part by grant no. NSF PHY-2309135 to the Kavli Institute for Theoretical Physics (KITP).
  L.M.B. and D.C. are partially supported by NASA through grant 80NSSC24K0173 and NSF through grant AST-2510951.
  This research used resources of the Argonne Leadership Computing Facility, which is a U.S. DOE of Science User Facility operated under contract DE-AC02-06CH11357. A.V. is supported by Université PSL through the PSL Young Researcher Starting Grant 2025 (2025-396) and received funding under the French Government program “Investissements d’Avenir". 
  This project was provided with computing HPC and storage resources by GENCI at TGCC and CINES, thanks to the grants 2025-A0180415130 and 2026-A0200415130 on the supercomputer Joliot Curie and Adastra's ROME, V100, and MI250x partitions. H.H. would like to acknowledge the support from the NVIDIA Corporation Academic Hardware Grant Program. E.G. is supported by the NSF Cyberinfrastructure for Sustained Scientific Innovation grant 2311800.
\end{acknowledgements}

\begin{appendix}

  \section{High-order shape functions in curvilinear coordinates}
  \label{sec:curvilinear}

  Throghout this paper, the discussion and all the unit tests were performed in Cartesian coordinates. \entity~however, uses general curvilinear formulation of both the Maxwell's equations, the current deposition and the equations of particle motion. In particular, as noted by \cite{Hakobyan2025}, charge conservation in curvilinear coordinates can be reformulated as 
  
  \begin{equation}
    \label{eq:curvspace-chargecons}
    \frac{\partial \rho }{\partial t} + \frac{1}{\sqrt{h}}\partial_i(\sqrt{h}J^i)= \sum_s \sum_p w_p q_s \frac{1}{\sqrt{h(\bm{x}_p)}}\frac{\partial}{\partial t}S_\mathcal{O}(\bm{x}-\bm{x}_p)+\frac{1}{\sqrt{h}}\partial_i\mathcal{J}^i=0,
  \end{equation}

  \noindent where $J^i$ is the $i$-th contravariant component of the current density, $\mathcal{J}^i\equiv\sqrt{h}J^i$, $\partial_i\equiv\partial/\partial x^i$, and the Einstein's summation by indices is implied. $\sqrt{h}$ here represents the square-root of the determinant of the spatial part of the metric tensor, which in flat space-time effectively quantifies the volume of the cell. Defining the charge density of each particle with the volume of the cell the particle is contained in, $\sqrt{h(\bm{x}_p)}=\sqrt{h}$,\footnote{This is ultimately a matter of choice, which determines whether the size of the shape function changes depending on the position of the particle within the cell. The charge conservation itself does not rely on this choice, as the equation $\partial_t (\partial_i(\sqrt{h}E^i)) = -4\pi \partial_i (\sqrt{h}J^i)$ is automatically satisfied regardless of how the shape function is defined (as long as its definition is consistent).} we arrive at a simple expression on the so-called ``conformal'' currents:

  \begin{equation}
      \sum_s \sum_p w_p q_s \frac{\partial}{\partial t}S_\mathcal{O}(\bm{x}-\bm{x}_p)+\partial_i\mathcal{J}^i=0,
  \end{equation}

  \noindent the finite-difference expression of which is exactly the same as for equation~\eqref{eq:chargeCont}, and can thus be resolved linearly using the methods discussed in Sec.~\ref{sec:esirkepov}. The main difference is that after particles deposit the conformal currents, $\mathcal{J}^i$, the physical contravariant current densities need to be recovered locally before using them in the Amp\`ere's law: $J^i\equiv \mathcal{J}^i/\sqrt{h}$.

  \section{Shape functions}
  \label{sec:shapeFunctions}

The shape functions in \entity~are constructed from a B-spline, as in \citet{Derouillat2018}. We define the interpolation kernel as a gate function, $\Pi(x)$:
\begin{align}
     \mathcal{S}_0(x) \,=\, \Pi(x) \,=\,
    \begin{cases}
      1 \quad &\text{if } \: \vert x \vert < 1/2\,
      \\
      0  \quad & {\rm else}\,
    \end{cases}
\end{align}
where the index of $\mathcal{S}$ refers to the order of the polynomial. Higher orders are then obtained through successive convolution:
\begin{align}
     \mathcal{S}_{\mathcal{O}}(x) \,=\, (\,\Pi \,\underbrace{ \star \ldots \star}_{\mathcal{O} \,{\rm times}} \Pi \,) (x) ,
\end{align}

\noindent or explicitly

\begin{equation}
    \mathcal{S}_\mathcal{O}(x) = \frac{1}{\vert x\vert} \Pi(x) \otimes \mathcal{S}_{\mathcal{O}-1}(x) \equiv \frac{1}{\vert x\vert} \int\limits_{-\infty}^\infty dx' \: \Pi(x' - x) \mathcal{S}_{\mathcal{O}-1}(x') \: .
\end{equation}

\noindent Note that in our notation the order $\mathcal{O}$ refers to the leading order of the resulting polynomial, while in \citet{Derouillat2018} the 0\thu-order refers to a point-like quasi-particle, represented by a Dirac delta function.
Hence, to match the definitions their $s^{(1)}(x)$ is equivalent to our $\mathcal{S}_0(x)$.

\noindent Solving this convolution up-to 5\thu-order yields:

  \begin{align}
    \mathcal{S}_2(x) =
    \begin{cases}
      \frac{3}{4} - \vert x\vert ^2 \quad &\text{if } \: \vert x \vert < 1/2
      \\
      \frac{1}{2} \left( \frac{3}{2} - \vert x\vert \right)^2 \quad &\text{if } \: 1/2 \leq \vert x \vert < 3/2
      \\
      0  \quad&\text{if } \: \vert x \vert \geq 3/2
    \end{cases}
  \end{align}
  \begin{align}
    \mathcal{S}_3(x) =
    \begin{cases}
      \frac{1}{6} \left( 4 - 6 x^2 + 3 \vert x\vert ^3\right) \quad &\text{if } \: \vert x \vert < 1
      \\
      \frac{1}{6} \left( 2 - \vert x\vert \right)^3 \quad &\text{if } \: 1 \leq \vert x \vert < 2
      \\
      0  \quad&\text{if } \: \vert x \vert \geq 2
    \end{cases}
  \end{align}
  \begin{align}
    \mathcal{S}_4(x) =
    \begin{cases}
      \frac{115}{192} - \frac{5}{8}x^2 + \frac{1}{4} x^4 \quad &\text{if } \: \vert x \vert < 1/2
      \\
      \frac{55}{96} + \frac{5}{24}  \vert x  \vert - \frac{5}{4} x^2 + \frac{5}{6}  \vert x  \vert^3 + \frac{1}{6} x^4 \quad &\text{if } \: 1/2 \leq \vert x \vert < 3/2
      \\
      \frac{625}{384} - \frac{125}{48}  \vert x  \vert + \frac{25}{16} x^2 + \frac{5}{12}  \vert x \vert^3 + \frac{1}{24} x^4 \quad &\text{if } \: 3/2 \leq \vert x \vert < 5/2
      \\
      0  \quad&\text{if } \: \vert x \vert \geq 5/2
    \end{cases}
  \end{align}
  \begin{align}
    \mathcal{S}_5(x) =
    \begin{cases}
      \frac{11}{20} - \frac{1}{2}x^2 + \frac{1}{4} x^4 - \frac{1}{12}  \vert x \vert^5 \quad &\text{if } \: \vert x \vert < 1
      \\
      \frac{17}{40} + \frac{5}{8}  \vert x \vert - \frac{7}{4} x^2 + \frac{5}{4}  \vert x \vert^3 - \frac{3}{8} x^4 + \frac{1}{24}  \vert x \vert^5 \quad &\text{if } \: 1 \leq \vert x \vert < 2
      \\
      \frac{81}{40} - \frac{27}{8}  \vert x \vert + \frac{9}{4} x^2 - \frac{3}{4}  \vert x \vert^3 + \frac{1}{8} x^4 - \frac{1}{120}  \vert x \vert^5 \quad &\text{if } \: 2 \leq \vert x \vert < 3
      \\
      0  \quad&\text{if } \: \vert x \vert \geq 3
    \end{cases}
  \end{align}

\noindent We refer to the publicly available source code of \entity~for the implementation of 6\thu~to 11\thu-order.

\end{appendix}

\bibliography{references}
\bibliographystyle{aasjournalv7}

\end{document}